\documentclass[reprint,aps,sort&compress,amsmath,amssymb,superscriptaddress]{revtex4-1}
\usepackage{bm}
\usepackage{graphicx}
\usepackage{units}
\usepackage[english]{babel}
\usepackage[colorlinks, citecolor=blue, urlcolor=blue, linkcolor=red,breaklinks]{hyperref}

\begin{document}
	
	\title{Designing Moir\'e Patterns by Strain}
	
	\author{Federico Escudero}
	\email{federico.escudero@uns.edu.ar}
	\affiliation{IFISUR, Departamento de F\'isica, CONICET, Universidad Nacional del Sur, Av. Alem 1253, B8000 Bah\'ia Blanca, Argentina}
	\affiliation{IMDEA Nanoscience, Faraday 9, 28049 Madrid, Spain}
	\author{Andreas Sinner}
	\affiliation{Institute of Physics, University of Opole, 45-052 Opole, Poland}
	\author{Zhen Zhan}
	\affiliation{IMDEA Nanoscience, Faraday 9, 28049 Madrid, Spain}
	\author{Pierre A. Pantale\'on}
	\email{pierre.pantaleon@imdea.org}
	\affiliation{IMDEA Nanoscience, Faraday 9, 28049 Madrid, Spain}
	\author{Francisco Guinea}
	\affiliation{IMDEA Nanoscience, Faraday 9, 28049 Madrid, Spain}
	\affiliation{Donostia International Physics Center, Paseo Manuel de Lardiz\'abal 4, 20018 San Sebastián, Spain}
	\affiliation{ Ikerbasque, Basque Foundation for Science, 48009 Bilbao, Spain}
	
	\begin{abstract}
		
		Experiments conducted on two-dimensional twisted materials have revealed a plethora of moiré patterns with different forms and shapes. The formation of these patterns is usually attributed to the presence of small strains in the samples, which typically arise during their fabrication. In this work we find that the superlattice structure of such systems actually depends crucially on the interplay between twist and strain. For systems composed of honeycomb lattices, we show that this can lead to the formation of practically any moiré geometry, even if each lattice is only slightly distorted. As a result, we show that under strain the moiré Brillouin zone is not a stretched irregular hexagon, but rather a primitive cell that changes according to the geometry of the strained moiré vectors. We identify the conditions for the formation of hexagonal moiré patterns arising solely due to shear or biaxial strain, thus opening the possibility of engineering moiré patterns solely by strain.
		Moreover, we study the electronic properties in such moiré patterns and find that the strain tends to suppress the formation of the flat moiré bands, even in the strain-induced hexagonal patterns analogous to those obtained by the twist only. Our work explains the plethora of moiré patterns observed in experiments, and provides a solid theoretical foundation from which one can design moiré patterns by strain.
		
	\end{abstract}
	
	\maketitle  
	
	\section{Introduction}\label{sec:Introduction}
	
	The recent discovery of correlated electronic states and superconductivity in twisted bilayer graphene (TBG)\cite{Andrei2021Moire,Jarillo2018a,Jarillo2018b} has sparked a great interest in twisted moir\'e systems. Theoretical works in TBG~\cite{LopesdosSantos2007,Shallcross2008TwistB, Shallcross2010Turbo,Bistritzer2011,TramblydeLaissardire2010,Mele2010Comm, LopesdosSantos2012}, and transition metal dicalcogenides (TMDs)~\cite{Xian2021MoS,Angeli2021Gamma,Naik2018UltraFlat,Wu2018TMD,Tang2020Hubb,Regan2020Mott,Wang2020Corre,Ni2019Soliton,Xian2019Multi,Koegl2023}, have demonstrated that the moir\'e patterns in these systems can give rise to narrow bands that are largely responsible for the correlated effects~\cite{Jarillo2018a,Jarillo2018b,yankowitz2019,lu2019superconductors,polshyn2019large,sharpe2019emergent,serlin2020intrinsic,chen2020tunable,saito2020independent,zondiner2020cascade,wong2020cascade,stepanov2020untying,xu2020correlated,choi2021correlation,rozen2021entropic,cao2021nematicity,stepanov2021competing,oh2021evidence,xie2021fractional,berdyugin2022out,turkel2022orderly,huang2022observation}. The form of these moir\'e patterns, however, can be highly sensitive to the presence of strain in the system~\cite{Ketal21}. This can have significant effects on the electronic properties, e.g., by preventing the bands from becoming flat around the magic angle, or by splitting the van Hove singularities~\cite{Huder2018Hetero,Fu2019,Mannai2021,Metal21,Wang2023Open}. Although the strain in superlattices configurations typically arises randomly during the fabrication of the samples~\cite{Hsieh2023, Cazeaux2022}, recent experimental advances have opened the possibility of inducing and controlling, in a precise way, different types of strain fields~\cite{Pea2023}. This provides a promising platform for designing moir\'e patterns, and tune the electronic properties, through the interplay between twist and strain~\cite{Ketal22b}.
	
	In superlattice configurations, the effect of the strain is usually magnified in the resulting moir\'e pattern~\cite{DWCF14}. Local variations of strain in the samples can, indeed, lead to large changes in the moir\'e pattern~\cite{Engelke2022Strain}. This is consistent with several recent experimental studies where the creation of different types of moir\'e lattice defects have been reported. Examples include domain walls between different stacking domains in TBG~\cite{Alden2013Soliton}, hexagonal boron nitride~\cite{Woods2021}, or TMDs~\cite{Shabani2021}. On the other hand, the effect of strains in monolayer graphene and other non-twisted bidimensional materials has been extensively studied~\cite{Pozrikidis_2008,VKG2010,DWCF14,Aetal16,Naumis2017Review}, and important insights on the role of strains in twisted bilayer graphene have been described in~\cite{Huder2018Hetero,Qiao2018, Fu2019,Engelke2022Strain,Cazeaux2022,Ahmed2023Mesoscale}. Interestingly, highly anisotropic moir\'e patterns in strained twisted bilayer graphene have been reported in many experiments~\cite{Alden2013Soliton,Qiao2018,Woods2021,Mendoza2021,Shabani2021,Ketal21,Jetal22,Jetal22b,Ketal22c,Andrei2022}. In addition to anisotropies, almost every experiment in multilayer graphene~\cite{Alden2013Soliton,Jiang2017,Mendoza2021,Craig2023tr,Engelke2022Strain,Cazeaux2022,Hesp2023} and TMDs~\cite{Bai2020,Shabani2021,Andrei2022}, have shown the existence of moir\'e patterns with different geometries. In particular, recent experiments have shown the existence of unconventional rectangular moire patterns in TMDs~\cite{Andrei2022} and multilayer graphene~\cite{Hesp2023}. 
	
	Inspired by these findings, in this work we study how the interplay between twist and strain can modify the geometrical properties of the moir\'e patterns. We find that by selectively applying strain to the system one can  change the moir\'e patterns to practically any geometry, even at very small strain magnitudes that only slightly distort each lattice. Exploiting a unique transformation that determines the relative angle and length between the moir\'e vectors, we develop a general theoretical scheme which allows one to describe any strained moir\'e geometry. We discuss different experimentally relevant types of strain, such as uniaxial heterostrain, shear strain and biaxial strain. We obtain and discuss the formation of special moir\'e geometries, such as the square moire patterns. We also show that hexagonal moir\'e patterns, analogous to those obtained with only a twist, can be formed solely by the application of shear or biaxial strain, thus opening the possibility of engineering moir\'e patterns only by strain. Finally, we observe that the typical irregular hexagonal cell, commonly used to describe strained honeycomb lattices, is no longer the moiré Brillouin zone (mBZ) of the strained superlattice. Instead, we identify a family of mBZ, with distinct geometries, that reflect the symmetries of the superlattice.
	
	Our geometrical analysis of strained moiré patterns overlaps with those recently presented in Ref.~\cite{Koegl2023}, where various types of strain have also been examined. However, despite the similarities, our theoretical scheme is build upon finding a unique transformation that directly determines the geometrical properties of the moiré vectors. This allows us to analytically study, in greater detail, what combinations of twist and strain result in any particular moiré pattern, thus providing a firm platform from which one can actually design moiré patterns. In addition, we develop a comprehensive account of the strain effects in both real and reciprocal space, and in particular discuss how these can strongly reshape the moiré BZ, which has not been addressed before in the literature. We thus believe that our work complements previous theoretical studies by providing a detailed account of how the moiré patterns can be actually designed by strain.
	
	Furthermore, our geometrical analysis is complemented by the studies of the electronic properties. We find that the modification of the moir\'e patterns by the strain plays a crucial role in the formation of flat bands around a magic angle. We attribute this to an interplay between the shift of the Dirac points in each deformed lattice, due to geometric and energetic effects, and the moir\'e potential that couples them. The strain influences both by breaking almost all the symmetries in the system, effectively preventing the lowest moir\'e bands to flatten across the BZ. We find that this occurs even in the hexagonal moir\'e patterns that arise due to strain only, and that on the moir\'e scale look practically identical to those obtained with the twist only. In these cases the strain reorganizes the charge density in the system, and leads to the splitting and appearance of multiple high-order van Hove singularities. 
	
	The rest of the paper is organized as follows: In Section \ref{sec:geometry} we discuss geometrical properties of strained moiré patterns. We describe in details how different patterns can be achieved through the interplay between twist and different types of strain. We also obtain how the first moiré Brillouin zone changes under strain. In Section \ref{sec:ElectronicProperties} we discuss the electronic properties of the strained moiré patterns, using an extension of the continuum model in the presence of strain. We calculate the band structure, the density of states, and the charge density profile under different types of strain, and compare them to the case of TBG without strain. Finally, our conclusions follow in Section \ref{sec:conclusions}.
	
	\medskip
	
	\section{Geometrical properties of strained moir\'e patterns}\label{sec:geometry}
	
	\subsection{General considerations}\label{subsec:GeneralStrain}
	
	We choose the lattice vectors of a honeycomb lattice as $\mathbf{a}_{1}=a\left(1,0\right)$ and $\mathbf{a}_{2}=a\left(1/2,\sqrt{3}/2\right)$, where $a$ is the lattice constant ($a\simeq2.46\:\textrm{Å}$ in graphene). In a honeycomb twisted bilayer configuration, the usual rotation by $\pm\theta/2$, and a further application of strain, yields $\tilde{\mathbf{a}}_{i,\pm}=\left(\mathbb{I}+\mathcal{E}_{\pm}\right)\mathrm{R}\left(\pm\theta/2\right)\mathbf{a}_{i}$, where $\mathbb{I}$ is the $2\times2$ identity matrix, $\mathrm{R}\left(\theta\right)$ is the rotation matrix and $\mathcal{E}_{\pm}$ is the strain tensor. At small deformations (that is, to leading order in $\mathcal{E}_{\pm}$), the reciprocal vectors can be obtained as $\tilde{\mathbf{b}}_{i,\pm}\simeq\left(\mathbb{I}-\mathcal{E}_{\pm}\right)\mathrm{R}\left(\pm\theta/2\right)\mathbf{b}_{i}$. In what follows, we only restrict our discussion to small twist angles and the practical case in which the forces act oppositely in each layer, $\mathcal{E}_{+}=-\mathcal{E}_{-}=\mathcal{E}/2$. Then, for a general strain tensor of the form $\mathcal{E}=\sum_{ij}\epsilon_{ij}\left(\mathbf{e}_{i}\otimes\mathbf{e}_{j}\right)$ (where $i,j=x,y$), the moir\'e lattice vectors can be obtained as $\mathbf{g}_{i}=\tilde{\mathbf{b}}_{-,i}-\tilde{\mathbf{b}}_{+,i}$, which implies $\mathbf{g}_{i}=\mathbf{T}\mathbf{b}_{i}$, where
	\begin{align}
		\mathbf{T} & =\left(\mathbb{I}+\mathcal{E}/2\right)\mathrm{R}\left(-\theta/2\right)-\left(\mathbb{I}-\mathcal{E}/2\right)\mathrm{R}\left(\theta/2\right).
		\label{eq:T}
	\end{align} 
	We are interested in how the combination of rotation and strain changes the geometry of the moir\'e patterns. The angle $\beta$ between the moir\'e vectors can be determined from the symmetric transformation $\mathbf{F}=\mathbf{T}^{\mathrm{T}}\mathbf{T}$ acting on the reciprocal vectors $\mathbf{b}_{i}$,
	\begin{equation}
		\cos\beta=\frac{\mathbf{F}\mathbf{b}_{1}\cdot\mathbf{b}_{2}}{\sqrt{\left(\mathbf{F}\mathbf{b}_{1}\cdot\mathbf{b}_{1}\right)\left(\mathbf{F}\mathbf{b}_{2}\cdot\mathbf{b}_{2}\right)}}.\label{eq:cosb}
	\end{equation}
	We can separate $\mathbf{F}=\mathbf{F}_{0}+\mathbf{F}_{\epsilon}$, where $\mathbf{F}_{0}$ is the contribution due to pure rotations, and $\mathbf{F}_{\epsilon}$ is the contribution due to the combination of rotation and strain:
	\begin{align}
		\mathbf{F}_{0} & =4\sin^{2}\left(\theta/2\right)\mathbb{I},\label{eq:F0}\\
		\mathbf{F}_{\epsilon} & =\sin\theta\left(\begin{array}{cc}
			-2\epsilon_{xy} & \epsilon_{xx}-\epsilon_{yy}\\
			\epsilon_{xx}-\epsilon_{yy} & 2\epsilon_{xy}
		\end{array}\right)+\cos^{2}\left(\theta/2\right)\mathcal{E}^{2}.\label{eq:Fe}
	\end{align}
	Since $\mathbf{F}_{0}$ is a spherical tensor, a transformation by $\mathbf{F}_{0}$ alone does not change $\beta$. This is, of course, the situation without strain, where the honeycomb layers are only rotated and the moir\'e vectors have always the same angle $\beta=2\pi/3$. However, under strain the vectors are also transformed by the non-spherical tensor $\mathbf{F}_{\epsilon}$, which changes the angle of $\mathbf{b}_{i}$ and hence modifies the geometrical properties of the moir\'e pattern. Note that the second term in Eq.~\eqref{eq:Fe} describes the possibility of obtaining moir\'e patterns without rotations, i.e., purely by strain~\cite{SanJose2012Gauge,Georgoulea2022}. 
	
	Equations (\ref{eq:T}) to (\ref{eq:Fe}) constitute the central results of the geometrical part of our study. They possess the versatility to describe a wide range of moir\'e structures, relying solely on the transformation matrix $\mathbf{F}$. This matrix can be constructed using an arbitrary strain tensor, rotation matrix, and even lattice geometries with appropriately chosen lattice vectors. These equations provide a concise and straightforward representation, that can also be employed to reproduce the results presented in Ref.~\cite{Koegl2023}.
	
	One crucial aspect of the modification of moir\'e patterns under strain is that it requires significantly smaller strain magnitudes compared to the strain needed to modify a monolayer honeycomb lattice. This can be observed by examining the strain required to change the angle between the corresponding lattice vectors. Consider, for instance, the case of uniaxial heterostrain along the $\phi=0$ direction, i.e., $\epsilon_{xx}=\epsilon$, $\epsilon_{yy}=-\nu\epsilon$ and $\epsilon_{xy}=0$, where $\nu$ is the Poisson ratio. As in ~Eq. \eqref{eq:cosb}, we can obtain the angle $\alpha$ between the strained reciprocal vectors $\tilde{\mathbf{b}}_{\pm}$ through the symmetrical transformation $\mathbf{T}_{\pm}^{\mathrm{T}}\mathbf{T}_{\pm}$, where $\mathbf{T}_{\pm}=\left(\mathbb{I}\mp\mathcal{E}/2\right)\mathrm{R}\left(\pm\theta/2\right)$. 
	Then, at low twist angle and to leading order in $\epsilon$ we get
	\begin{align}
		\cos\alpha & \simeq-\frac{1}{2}\mp\frac{3}{16}\epsilon\left(\nu+1\right),\\
		\cos\beta & \simeq-\frac{1}{2}+\frac{3}{16}\epsilon\left(\nu+1\right)\frac{2\sqrt{3}}{\theta}.
	\end{align}
	Thus at small values of $\theta$ one needs much smaller strain magnitudes to modify $\beta$ than to modify $\alpha$. In fact, for experimentally relevant values $\epsilon\lesssim10\%$, and sufficiently low twist angles~\cite{Bai2020,Ketal21,Pea2023}, one can, in principle, vary the angle $\beta$ to any value between $0$ and $\pi$. In comparison, for the same strain range, the actual angle between the lattice vectors in the monolayer varies only by just a few degrees~\cite{Pereira2009uni} (see also~\cite{Naumis2017Review} and references therein). This means that, under the right strain parameters, the moir\'e patterns can be practically changed to any desired geometry, even if the underlying honeycomb lattices are only slightly distorted \cite{Hesp2023,Koegl2023}. Such behavior is possible because the moir\'e pattern arises from the twist angle or the lattice mismatch between the two monolayers, and any small distortion is enhanced by the moir\'e~\cite{DWCF14}. 
	
	\begin{figure}[t]
		\includegraphics[scale=0.14]{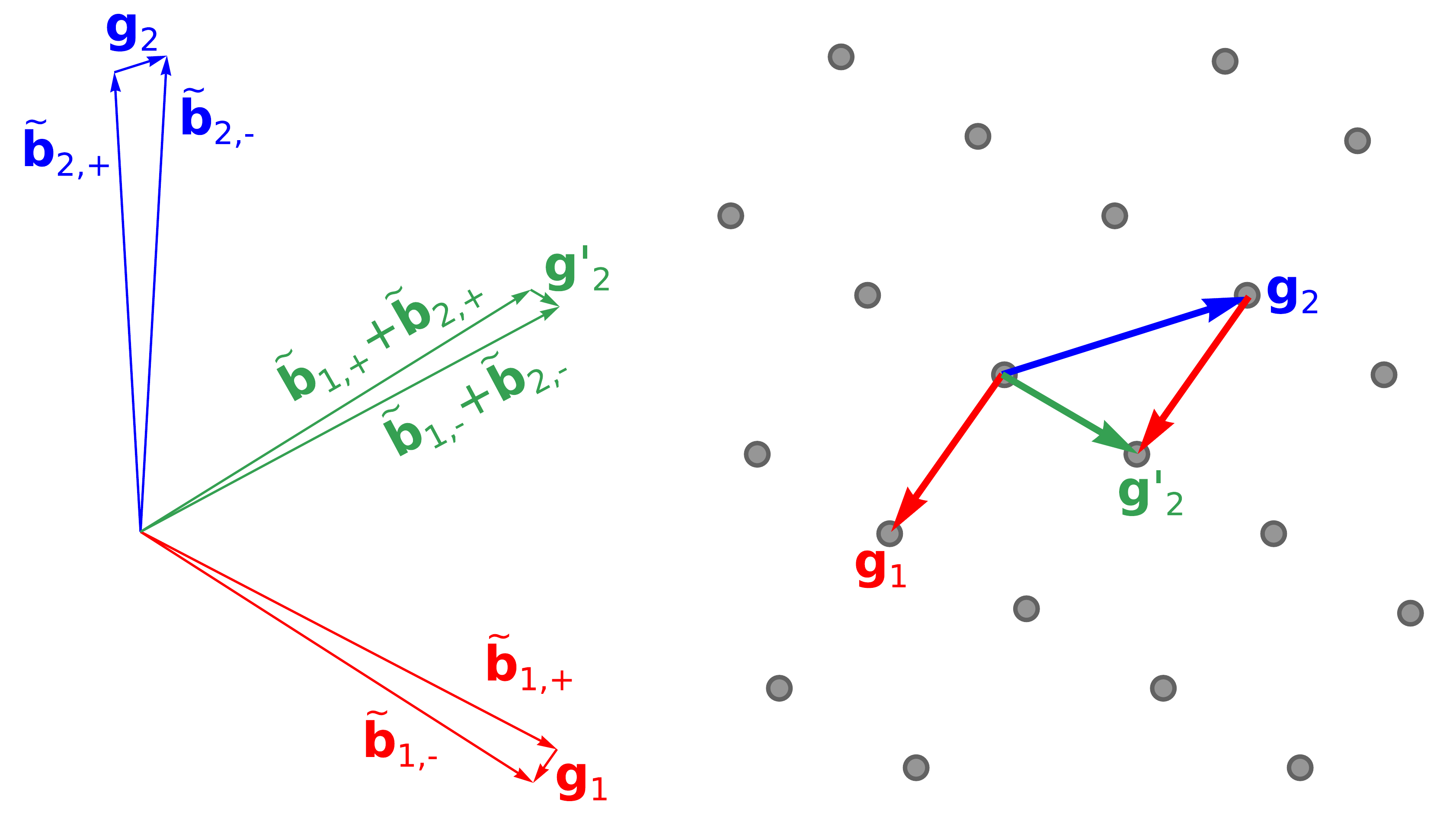}	
		\caption{Construction of the  moir\'e vectors for the case of uniaxial heterostrain with $\theta=5^{\circ}$, $\epsilon=5\%$ and $\phi=60^{\circ}$. The strained reciprocal lattice vectors $\tilde{\mathbf{b}}_{1,\pm}$ and $\tilde{\mathbf{b}}_{2,\pm}$ in top and bottom layer are shown in red and blue, from which the  moir\'e vectors $\mathbf{g}_{i}=\tilde{\mathbf{b}}_{i,-}-\tilde{\mathbf{b}}_{i,+}$ are obtained. The superlattice spanned by these vectors is shown on the right. In this case the vector $\mathbf{g}_{2}$ is not the shortest one that can be taken, since it can be translated by $\mathbf{g}_{1}$ to obtain a shorter  moir\'e vector $\mathbf{g}'_{2}=\mathbf{g}_{2}+\mathbf{g}_{1}$. The superlattice vector $\mathbf{g}'_{2}$ arises from constructing the second  moir\'e vector by taking the difference between the reciprocal vectors $\left(\tilde{\mathbf{b}}_{1,\pm}+\tilde{\mathbf{b}}_{2,\pm}\right)$.}\label{fig:mvectors}
	\end{figure}
	
	\begin{figure*}[t]
		\includegraphics[scale=0.36]{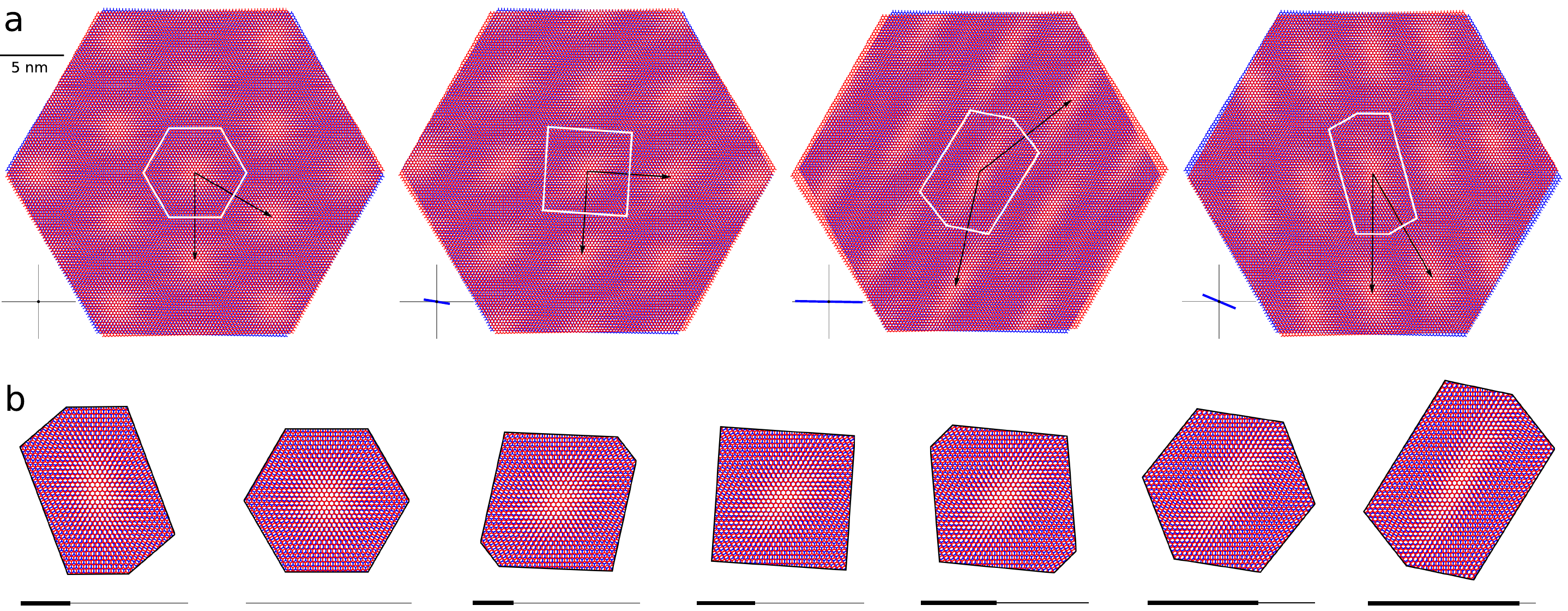}	
		\caption{Moir\'e patterns generated by uniaxial heterostrain in a twisted honeycomb lattice with $\theta=2{^\circ}$, for the case of equal length moir\'e vectors. Panel (a):  Strain parameters vary from left to right as: $\epsilon=0$ (no strain); $\epsilon\simeq1.64\%,\phi\simeq-9.40{^{\circ}}$; $\ensuremath{\epsilon\simeq4.44\%,\phi\simeq-0.90{^{\circ}}}$; and $\epsilon\simeq-2.28\%,\phi\simeq-22.7{^{\circ}}$. The corresponding angles between the primitive moir\'e vectors, shown in black, are $\beta_{R}=60{^\circ},90{^\circ},140{^\circ},30{^\circ}$. The Wigner--Seitz cells of the moir\'e superlattices are shown in white. The insets underneath each panel (in blue) show the strain magnitude in a scale of $5\%$, and the strain angle relative to the non-rotated lattice orientation. Panel (b): The evolution of the moir\'e pattern within the Wigner--Seitz cell for angles (from left to right)  $\beta_{R}=40{^\circ},60{^\circ},80{^\circ},90{^\circ},100{^\circ},120{^\circ},140{^\circ}$. Bars underneath (thick line) indicate the magnitude of the strain $\epsilon$ in a scale of $5\%$ (thin line). As $\epsilon$ increases, the stretch of the AA stacking within the primitive cell increases. As a result, the shape of two moir\'e patterns with the same periodicity is not the same if they correspond to different strain magnitudes.}
		\label{fig:MoireFamily}
	\end{figure*}
	
	It is important to note that, under strain, the moir\'e vectors obtained by using a unique construction (e.g., by the difference between $\tilde{\mathbf{b}}_{-,i}$ and $\tilde{\mathbf{b}}_{+,i}$) may not be the smallest ones (the so called \textit{primitives}). Consequently, for arbitrary strain parameters the moir\'e vectors may not reflect the symmetries of the corresponding moir\'e geometry. For example, a square moir\'e pattern can result from equal length moir\'e vectors with an angle $\beta=\pi/2$ between them, but also from moir\'e vectors with an angle $\beta=\pi/4$ and relative length $\left|\mathbf{g}_{1}\right|/\left|\ \mathbf{g}_{2}\right|=\sqrt{2}$. In general, for arbitrary strain parameters, the primitive moir\'e vectors are obtained by appropriately changing the set of reciprocal lattice vectors from which they are constructed (e.g., by taking the difference between the deformed vectors $\mathbf{b}_{1}$ and $\mathbf{b}_{1}+\mathbf{b}_{2}$, rather than $\mathbf{b}_{1}$ and $\mathbf{b}_{2}$; see Fig. \ref{fig:mvectors}). These different constructions of $\mathbf{g}_{i}$ eventually reflect the underlying symmetries of the honeycomb lattices.
	
	Furthermore, although the form of the  moir\'e patterns between nonstrained honeycomb lattices are uniquely determined by the \textit{periodicity} of the superlattice, the same is not true under strain. Two set of symmetric moir\'e vectors, which technically describe the same superlattice, may actually correspond to different forms of the moir\'e pattern. The reason is that the \emph{stretch} of the AA stacking, which is \textit{periodically} repeated by the moir\'e vectors, increases under strain~\cite{Ketal22b,Mendoza2021, Hesp2023}. The effect of the strain on the moir\'e patterns thus acts not only on the modification of the moir\'e periodicity, but also on the form of the stacking shape that is repeated. This behavior is, in a way, similar to the usual description of crystal structures through a basis within a primitive cell and a set of Bravais vectors that repeat such a basis. 
	
	Throughout this work we shall consider twist and strain parameters for which the resulting superlattice is not, in general, commensurate. However, recent experiments have demonstrated that even when the superlattice structure is not commensurate, there is a self-alignment due to lattice relaxation~\cite{Lai2023Imaging}. Simple commensurate cases are usually only feasible in special scenarios, such as in moiré patterns arising from only a twist~\cite{Mele2010Comm,Shallcross2010Turbo}, or from only certain types of strain~\cite{Shin2021,Long2022Atomistic}. Nevertheless, the geometrical properties of the system can still be well described through the analysis of the moiré vectors, since these can always be defined, regardless of whether the superstructure is commensurate or not~\cite{Moon2014Optical}. 
	Furthermore, at low twist angles and strain magnitudes the moiré length is usually much larger than the atomic length, and the electronic properties of the system can be well described by effective continuum models (Sec. \ref{sec:ElectronicProperties}), even if the moiré pattern is incommensurate \cite{Koshino2015}.	
	
	Given the reciprocal vectors $\mathbf{g}_{i}$, the primitive moir\'e lattice vectors $\mathbf{g}_{i}^{R}$ are most easily obtained by the relation $\mathbf{g}_{i}\cdot\mathbf{g}_{j}^{R}=2\pi\delta_{ij}$, which implies $\mathbf{g}_{i}^{R}=\mathbf{T}^{-\mathrm{T}}\mathbf{a}_{i}$, where $\mathbf{T}$ is given by Eq. \eqref{eq:T}. Thus, the geometrical properties of the primitive moir\'e vectors are determined by the inverse transformation $\mathbf{F}^{-\mathrm{T}}=\mathbf{F}^{-1}$. In particular, the angle between the primitive vectors is $\beta_{R}=\pi-\beta$, where $\beta$ is the angle in reciprocal space given by Eq. \eqref{eq:cosb}. 
	
	In what follows we discuss in details the geometrical properties of the moir\'e patterns under three important kinds of strain: the uniaxial heterostrain, the biaxial strain, and the shear strain. It is worth mentioning that our formalism, applied here to moiré heterostructures which arise from honeycomb lattices, can be directly extended to other geometries by appropriately modifying the lattice vectors and the strain tensor \cite{Liu2018Tailoring}.
	
	\subsection{Uniaxial heterostrain}\label{subsec:Uniaxial} 
	
	The uniaxial heterostrain refers to a type of strain that is applied along a unique axis, and acts oppositely in each honeycomb lattice. From the experimental point of view it is widely regarded as the most relevant kind of strain in TBG. It was first introduced both theoretically and experimentally in Ref.~\cite{Huder2018Hetero}, and then further investigated in Refs.~\cite{Qiao2018,Fu2019,Metal21,Gao2021Hetero,Ketal21}. The developed approach can, nevertheless, be directly extended to other types of strains, as discussed in the next sections.
	
	The strain tensor of uniaxial heterostrain with magnitude $\epsilon$, along an angle $\phi$ relative to the $x$ axis, reads
	\begin{align}
		\mathcal{E} & =\mathrm{R}^{\mathrm{T}}\left(-\phi\right)\left(\begin{array}{cc}
			\epsilon & 0\\
			0 & -\nu\epsilon
		\end{array}\right)\mathrm{R}\left(-\phi\right)\nonumber \\
		& =\epsilon\left[\begin{array}{cc}
			\cos^{2}\phi-\nu\sin^{2}\phi & \left(1+\nu\right)\sin\phi\cos\phi\\
			\left(1+\nu\right)\sin\phi\cos\phi & \sin^{2}\phi-\nu\cos^{2}\phi
		\end{array}\right].
		\label{eq:Euniaxial}
	\end{align}
	The transformation matrix $\mathbf{F}$ then becomes
	\begin{align}
		\mathbf{F} & =4\sin^{2}\left(\frac{\theta}{2}\right)\mathbb{I}+\epsilon\left(1+\nu\right)\sin\left(\theta\right)\mathrm{R}\left(2\phi\right)\sigma_{x}\nonumber \\
		& +\cos^{2}\left(\frac{\theta}{2}\right)\frac{\epsilon^{2}}{2}\left[\left(1+\nu^{2}\right)\mathbb{I}+\left(1-\nu^{2}\right)\mathrm{R}\left(2\phi\right)\sigma_{z}\right],
		\label{eq:Funiaxial}
	\end{align}
	where $\sigma_{i}$ are the Pauli matrices.
	From here one can readily see that the solutions of Eq.~\eqref{eq:cosb} for the strain magnitude always scale with the twist angle as $\sim\tan\left(\theta/2\right)$. Indeed, by writing $\epsilon=\epsilon'\tan\left(\theta/2\right)$ it follows that $\mathbf{F}\propto\sin^{2}\left(\theta/2\right)$ and, consequently, that the angle equation for $\beta$ as function of $\epsilon'$ is independent of the twist angle. Thus, for any $\beta$ and $\phi$, the solutions of Eq.~\eqref{eq:cosb} for the strain magnitude have the form $\epsilon\propto\tan\left(\theta/2\right)$. What this general result reflects is that the lower the twist angle, the weaker strain is needed to modify the geometry of the  moir\'e superlattices. 
	
	\subsubsection{Equal length moir\'e vectors}
	
	In the following, to simplify our analysis, we focus on the moir\'e patterns formed by equal-length moir\'e vectors, i.e., on the structures with $\left|\mathbf{g}_{1}\right|=\left|\mathbf{g}_{2}\right|$. This choice allows for analytical solutions, which can be used to analyze the geometrical effects. The consideration of moir\'e vectors with different lengths is a straightforward extension of our analysis, as described in a following section. From Eq.~\eqref{eq:Funiaxial}, the equal-length moir\'e vector condition for non-zero strain is given by 
	\begin{equation}
		\epsilon_{\mathrm{eq}}=\frac{4}{1-\nu}\cot\left(\frac{\pi}{3}-2\phi\right)\tan\left(\frac{\theta}{2}\right).\label{eq:Eeq}
	\end{equation}
	Since $\epsilon_{\mathrm{eq}}\propto\tan\left(\theta/2\right)$ and thus $\mathbf{F}\propto\sin^{2}\left(\theta/2\right)$, Eq.~\eqref{eq:cosb} for $\epsilon=\epsilon_{\mathrm{eq}}$ does not depend on $\theta$. This is a rather remarkable result: it means that the strain direction $\phi$ needed to obtain the equal length  moir\'e vectors, with an angle $\beta$ between them, is independent of the twist angle. The twist angle only modifies the needed strain magnitude, and the resulting (equal) length of the  moir\'e vectors, $\left|\mathbf{g}_{i}\right|^{2}=\mathbf{F}\mathbf{b}_{i}\cdot\mathbf{b}_{i}\propto\sin^{2}\left(\theta/2\right)$ (as in the unstrained case). Note that Eq. \eqref{eq:Eeq} is not invariant under the transformation $\phi\rightarrow\phi+\pi/3$ because we are not considering the other solutions that are obtained by appropriately changing the construction of the moir\'e vectors (see Appendix \ref{App: HexagonalS}).
	
	As detailed in Appendix~\ref{App: Equal}, by solving Eq.~\eqref{eq:cosb} for $\phi$ one can get the required strain parameters in order to obtain the equal length moir\'e vectors with the angle $\beta$ between them. At low twist angles, the corresponding strain magnitudes are relatively small and well within the experimental range~\cite{Pea2023}. Some moir\'e patterns that can be formed under uniaxial heterostrain are shown in Fig. \ref{fig:MoireFamily}. In general, the moir\'e patterns are quite sensitive to the values of the strain parameters, in the sense that small changes in $\epsilon$ and $\phi$ can, in contrast, result in significant changes in the geometry of the moir\'e vectors~\cite{Alden2013Soliton,Engelke2022Strain}. Thus, the precise control over the magnitude and direction of the applied uniaxial heterostrain is crucial for designing  moir\'e patterns through the strain manipulation. It is worth noting that this control has already been achieved experimentally. Reference~\cite{Pea2023} describes a methodology for process-induced strain engineering, where the strain magnitude and direction in TBG can be controlled.
	
	Fig. \ref{fig:MoireFamily} also shows that the orientation of the Wigner--Seitz cell, and of the stretched AA stacking within it, change depending on the strain magnitude. This is because the strain modifies not only the angle between the moir\'e vectors, but also their orientation with respect to the non-strain case. For instance, in the strained case with $\beta_{R}=120^{\circ}$, the hexagonal primitive cell is rotated with respect to the same cell in the non-strain case. In general, the stretch of the AA stacking occurs along the direction of the moir\'e vector $\mathbf{g}_{1}^{R}\pm\mathbf{g}_{2}^{R}$, where $+$ ($-$) when $\beta_{R}<90^{\circ}$ ($\geq90^{\circ}$). Such direction always coincides with one corner of the Wigner--Seitz cell. The angle $\phi_{s}$ of the stretching can thus be estimated as
	\begin{equation}
		\cos\phi_{s}=\frac{\mathbf{g}_{1}^{R}\pm\mathbf{g}_{2}^{R}}{\left|\mathbf{g}_{1}^{R}\pm\mathbf{g}_{2}^{R}\right|}\cdot\mathbf{e}_{x}.\label{eq:stretchA}
	\end{equation}
	Note that $\phi_{s}$ generally differs from the strain angle $\phi$, i.e., the observed stretch of the AA stacking does not reflect the direction along which the uniaxial heterostrain is applied. It only reflects the magnitude of the applied strain.  Since when $\epsilon\propto\tan\left(\theta/2\right)$ one has $\mathbf{T}\propto\sin\left(\theta/2\right)$ [cf. Eqs. \eqref{eq:Euniaxial} and \eqref{eq:T}], it follows that for any strain direction $\phi$ that
	yields an angle $\beta_{R}$ between the moir\'e vectors, the corresponding stretch angle $\phi_{s}$ is independent of the twist angle $\theta$. The above analysis may allow one to estimate the strain properties of twisted bilayer honeycomb samples by analysing only the shape of the AA regions.
	
	\subsubsection{Special moir\'e patterns}
	
	\begin{figure}[t]
		\includegraphics[scale=0.35]{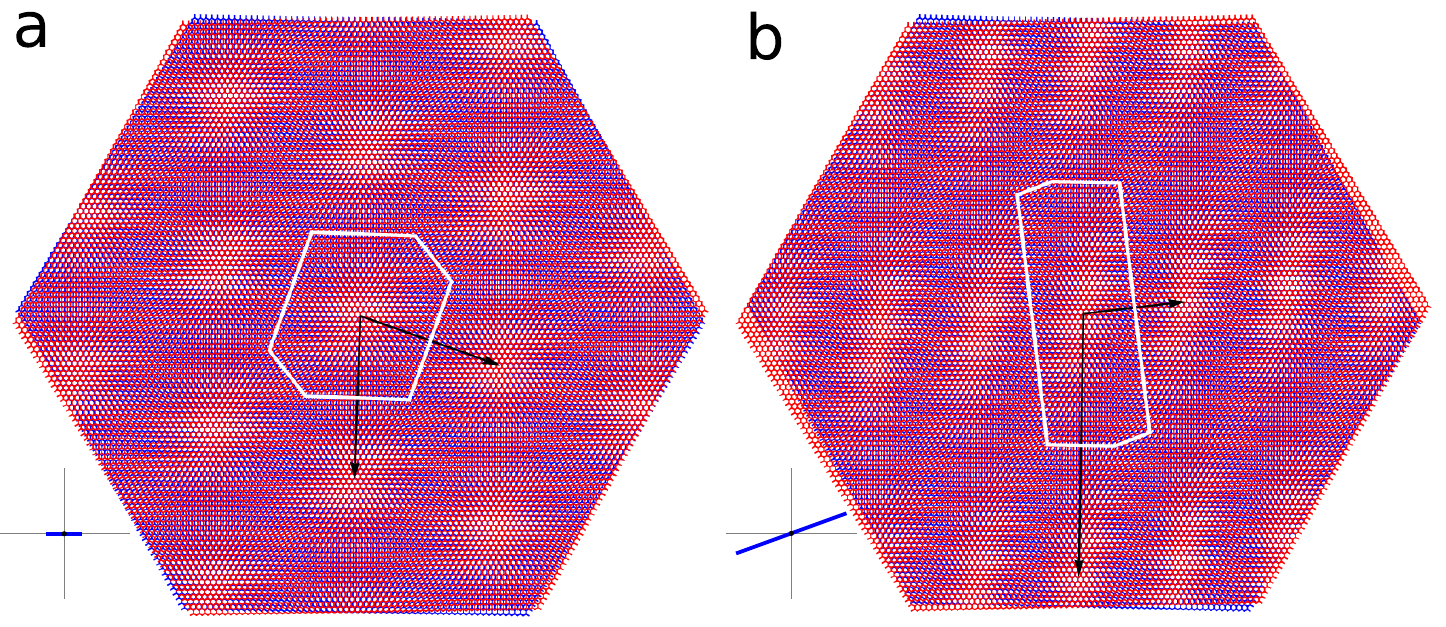}	
		\caption{Moir\'e patterns under uniaxial heterostrain with unequal length superlattice vectors, for twist angle $\theta=2{^\circ}$. The strain parameters are (a)  $\epsilon=1\%,\phi=0{^{\circ}}$	and (b) $\epsilon=3.5\%,\phi=20{^{\circ}}$. The corresponding relative length and angle between the moir\'e vectors are (a) $\left|\mathbf{g}_{1}^{R}\right|/\left|\mathbf{g}_{2}^{R}\right|\simeq1.11,\beta_R\simeq72.5^{\circ}$ and (b) $\left|\mathbf{g}_{1}^{R}\right|/\left|\mathbf{g}_{2}^{R}\right|\simeq2.57,\beta_R\simeq97.8^{\circ}$. The respective strain magnitude $\epsilon$ (in blue) and the strain angle direction $\phi$ relative to the non-rotated lattice orientation is shown in a scale of $4\%$ 
			underneath each panel.
		}\label{fig:moirenoneq}
	\end{figure}
	
	Some special moir\'e patterns that may be accomplished deserve further discussion. One case is the square moir\'e pattern shown in Fig.~\ref{fig:MoireFamily}a). Squared-like moir\'e patterns have already been experimentally observed~\cite{Jiang2017,Tilak2022Andrei,Yu2022}, and theoretically predicted~\cite{Koegl2023}. While their shape has been attributed to highly distorted  moir\'e patterns, our model indicates that this geometry can be alternatively obtained by the right combination of twist angle and strain. Another interesting case occurs when $\beta_R=120^{\circ}$, where one can have the same hexagonal moir\'e periodicity as with no strain (where $\beta_R=60^{\circ}$), albeit with a stretched AA stacking within the primitive cell (see Fig.~\ref{fig:MoireFamily}b). 
	
	A particularly relevant case is the critical limit in which the moir\'e vectors become collinear. This can lead to quasi-unidimensional channels that have been predicted~\cite{DWCF14,Sinner2023} and observed in several experiments~\cite{Alden2013Soliton,Woods2021,Mendoza2021,Shabani2021,Ketal21,Jetal22,Jetal22b,Ketal22c,Andrei2022,Hsieh2023, Craig2023tr, Pea2023,Engelke2022Strain,Bai2020,Liu2018Tailoring}. Plugging $\beta=\{0,\pi\}$ into Eq. \eqref{eq:betasolsE} yields a critical strain parameter~\cite{Sinner2023}
	\begin{equation}
		\epsilon_{c}=\pm\frac{2}{\sqrt{\nu}}\tan\left(\frac{\theta}{2}\right).\label{eq:scriticalU}
	\end{equation} 
	This expression for $\epsilon_{\mathrm{c}}$ is actually quite general, i.e. it always leads to the collinear moir\'e vectors, regardless of the strain angle $\phi$~\cite{Sinner2023,Koegl2023}. Technically, this is because at this critical strain the determinant of the matrix $\mathbf{F}$ vanishes, which means that it becomes non-invertible and the moir\'e vectors are no longer linearly independent. 
	
	\subsubsection{Arbitrary strain parameters}
	
	The general situation of arbitrary strain parameters (within the limit of small deformations) is, in many ways, qualitatively very similar to the special case of equal length moir\'e vectors. By fixing, for example, the strain angle to $\phi=0$, one can still obtain many different geometries in which the angle between the moir\'e vectors can be tuned solely by changing the strain magnitude. Examples of such moir\'e patterns are shown in Fig.~\ref{fig:moirenoneq}. There one sees that, although the length of one moir\'e vector may be more than double the length of the other one, the  moir\'e patterns follows a similar behavior to the simpler ones analyzed in Fig.~\ref{fig:MoireFamily}. Thus our discussion in the previous section is readily generalized to arbitrary strain parameters. In particular, the symmetry of the  moir\'e superlattice, and the magnitude of the strain, are always reflected in the shape of the Wigner--Seitz cell, and the stretch of the AA stacking within it. Furthermore, the direction of the AA stretching also follows, in general, the direction of the moir\'e vector $\mathbf{g}_{1}^{R}\pm\mathbf{g}_{2}^{R}$ [cf. Eq. \eqref{eq:stretchA}].
	
	In the case of pure uniaxial heterostrain, without a twist, the transformation given by Eq. \eqref{eq:Funiaxial} reduces to
	\begin{equation}
		\mathbf{F}=\frac{\epsilon^{2}}{2}\left[\left(1+\nu^{2}\right)\mathbb{I}+\left(1-\nu^{2}\right)\mathrm{R}\left(2\phi\right)\sigma_{z}\right].
	\end{equation}
	Since the second term is not a spherical tensor, the resulting moiré pattern is not hexagonal. 
	
	\subsection{Shear strain}\label{subsec:Shear} 
	
	\begin{figure*}[t]
		\includegraphics[width= \textwidth]{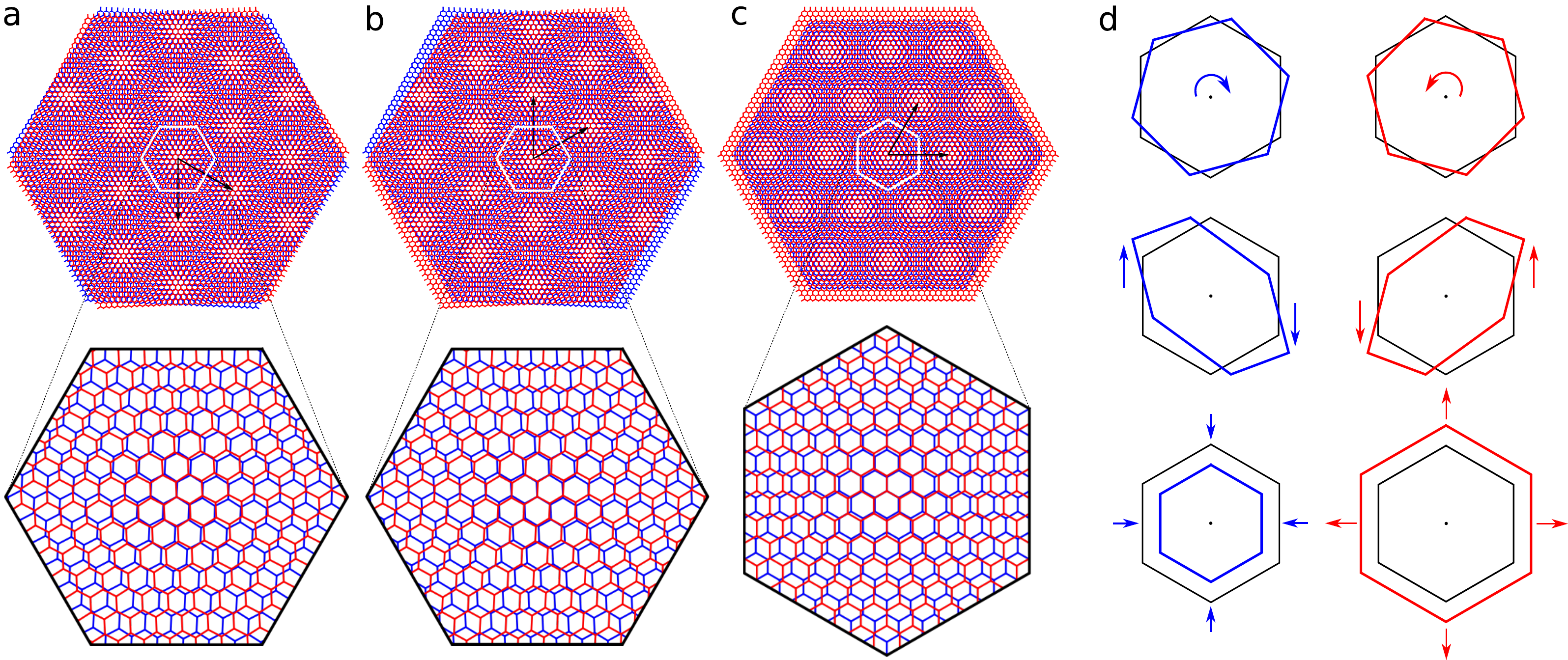}	
		\caption{Hexagonal moir\'e patterns generated only by: (a) twist angle $\theta=5^{\circ}$; (b) shear strain with magnitude $\epsilon_{s}=2\sin\left(\theta/2\right)\simeq8.7\%$; (c) biaxial strain with magnitude $\epsilon_{b}=2\sin\left(\theta/2\right)\simeq8.7\%$. From Eqs.~\eqref{eq:exytheta} and~\eqref{eq:bitheta}, all cases have the same moir\'e periodicity. The figures in the bottom row visualize the enlarged Wigner--Seitz cells. Here, the vicinity of the AA, AB and BA stacking positions looks different for each case. This difference, however, becomes smaller (and practically unnoticeable at the moir\'e scale) as the twist and strain decrease. Panel (d) shows schematically the corresponding deformations in the bottom (left) and top (right) lattices due to (from top to bottom panels) rotation, shear strain, and biaxial strain. The effects are exaggerated for better visualization.}
		\label{fig:TwistShearBiaxial}
	\end{figure*}
	
	In a honeycomb lattice, shear strain occurs when forces act parallel to its surface but in opposite directions. This leads to a distortion of the lattice. In simpler terms, shear strain in a honeycomb lattice comes from \textit{sliding} forces that deform the lattice without altering its overall volume, cf. Fig.~\ref{fig:TwistShearBiaxial}d). This kind of strain has been studied in graphene and transition metal dicalcohenides~\cite{Hovden2014Atomic, Liu2012Mechanical,Lee2017Strain,Wang2017Shear}. 
	
	The strain tensor due to shear forces applied perpendicularly to an angle direction $\varphi$ is given by
	\begin{align}
		\mathcal{E} & =\mathrm{R}^{\mathrm{T}}\left(-\varphi\right)\left(\begin{array}{cc}
			0 & \epsilon_{xy}\\
			\epsilon_{xy} & 0
		\end{array}\right)\mathrm{R}\left(-\varphi\right)\nonumber \\
		& =\epsilon_{s}\left(\begin{array}{cc}
			-\sin2\varphi & \cos2\varphi\\
			\cos2\varphi & \sin2\varphi
		\end{array}\right),
		\label{eq:shearS}
	\end{align}
	where $\epsilon_{xy}=\epsilon_{s}$ is the shear strain magnitude. For a twisted bilayer lattice, this leads to the transformation
	\begin{align}
		\mathbf{F} & =\left[4\sin^{2}\left(\frac{\theta}{2}\right)+\epsilon_{s}^{2}\cos^{2}\left(\frac{\theta}{2}\right)\right]\mathbb{I}\nonumber \\
		& \qquad-2\sin\left(\theta\right)\epsilon_{s}\mathrm{R}\left(2\varphi\right)\sigma_{z}.\label{eq:Fshear}
	\end{align}
	The second term implies that the combined effect of twist and shear strain can change the geometry of the moir\'e patterns, similar to the effect of uniaxial heterostrain. The main difference lies in how the distortion of each honeycomb lattice gives rise to a particular moir\'e geometry. Thus, although the moir\'e patterns for different strain types may appear similar, their electronic properties can be substantially different (cf. Fig.~\ref{fig:HexaBands}). 
	
	An interesting situation occurs in the case of pure shear forces without a twist angle, where Eq.~\eqref{eq:Fshear} reduces to $\mathbf{F}=\epsilon_{s}^{2}\mathbb{I}$. This transformation acts as in the twisted non-strain case, where $\mathbf{F}=4\sin^{2}\left(\theta/2\right)\mathbb{I}$,
	with the resulting moir\'e pattern being always hexagonal. This means that one can form hexagonal moir\'e patterns without any twist between the layers, just by applying opposite shear forces in each lattice, thus opening the possibility of engineering superlattice heterostructures purely by strain (cf. Fig~\ref{fig:TwistShearBiaxial}b). The shear angle $\varphi$ only changes the orientation of the moir\'e pattern. Interestingly, the moir\'e superlattice with pure shear strain can have the same periodicity as that of TBG with twist angle $\theta_{eq}$ if the strain magnitude satisfies  
	\begin{equation}
		\epsilon_{s}=2\sin\left(\frac{\theta_{eq}}{2}\right).
		\label{eq:exytheta}
	\end{equation}
	For example, a strain magnitude $\epsilon_{s}\sim1.8\%$ yields a moiré periodicity $L \sim 13.4$ nm, corresponding to an equivalent twist angle $\theta_{eq} \sim 1.05^{\circ}$.
	
	\subsection{Biaxial strain}\label{subsec:Biaxial} 
	
	In the case of biaxial strain the forces are equally applied along the $x$ and $y$ directions, and in opposite directions in each layer. The corresponding strain tensor reads $\mathcal{E}=\epsilon_b\mathbb{I}$, thus yielding the transformation matrix
	\begin{align}
		\mathbf{F} & =\left[4\sin^{2}\left(\frac{\theta}{2}\right)+\epsilon_{b}^{2}\cos^{2}\left(\frac{\theta}{2}\right)\right]\mathbb{I}.\label{eq:Fbiaxial}
	\end{align}
	Since $\mathbf{F}$ is always a spherical tensor, a biaxial strain cannot change the moir\'e geometry: any combination of strain and twist always results in a hexagonal moir\'e pattern. This is, of course, expected because the biaxial strain does not distort the hexagonal lattices, it only changes the size of the primitive cell. 
	The effect of twist and strain, in this case, is to only modify the orientation and length of the superlattice vectors. 
	
	The change of orientation can be measured in relation to the direction of the moir\'e vectors in the case of no strain, where, according to our reference convention, the second moir\'e vector in reciprocal space is always along the $x$ axis, $\mathbf{g}_{2}=8\pi\sin\left(\theta/2\right)/\sqrt{3}a\mathbf{e}_{x}$ [cf. Eq. \eqref{eq:T}]. In the case of biaxial strain, this moir\'e vector becomes $\mathbf{g}_{2}=\left(8\pi/\sqrt{3}a\right)\sin\left(\theta/2\right)\left[\mathbf{e}_{x}+\epsilon\cot\left(\theta/2\right)\mathbf{e}_{y}/2\right]$, so its angle $\alpha_{\epsilon}$ with respect to the $x$ axis reads
	\begin{equation}
		\cos\alpha_{\epsilon}=\frac{1}{\sqrt{1+\frac{\epsilon^{2}}{4}\cot^{2}\left(\theta/2\right)}}.\label{eq:alignB}
	\end{equation}
	
	By comparing Eqs. \eqref{eq:F0} and \eqref{eq:Fbiaxial} one can obtain the combinations of strain magnitude $\epsilon_b$ and twist angle $\theta$ that give the same moir\'e periodicity as with only a twist angle $\theta_{eq}$,
	\begin{equation}
		\sin^{2}\left(\theta/2\right)=\frac{\sin^{2}\left(\theta_{eq}/2\right)-\epsilon_b^{2}/4}{1-\epsilon_b^{2}/4}.\label{eq:EqualB}
	\end{equation}
	This condition does not, however, guarantee that both moir\'e patterns would be align, since their orientation may differ due to the strain effect. This can be important when one seeks an alignment between two (or more) moir\'e patterns arising from a combination of rotation
	and lattice mismatch. 
	
	A relevant example occurs in heterostructures of TBG/hBN in which hBN acts as a substrate of TBG~\cite{Yankowitz2012Emergence,Xue2011Scanning,Decker2011Local}. In this case, the lattice mismatch between graphene ($a_{g}=2.46\;\textrm{Å}$) and hBN ($a_{h}=2.50\;\textrm{Å}$)  can be accounted as a biaxial strain with magnitude $\epsilon_{b}\sim1-a_{T}/a_{B}=0.016$. If the twist angle in TBG is $\theta_T$, and the twist angle between hBN and the graphene layer directly on top is $\theta_{b}$, a \emph{moir\'e} \emph{alignment} implies that both moir\'e patterns have the same orientation and periodicity. Since in TBG the layers are only rotated, the orientation condition is obtained from Eq.~\eqref{eq:alignB} by setting $\cos\alpha_{\epsilon}=\pm1/2$, which gives $\theta_{b}\simeq\epsilon_{b}/\sqrt{3}\sim0.53^{\circ}$. Then the equal periodicity condition,~Eq.~\eqref{eq:EqualB}, implies that the twist angle in TBG should be $\theta_{T}\simeq\sqrt{\theta_{b}^{2}+\epsilon_{b}^{2}}\sim1.06^{\circ}$, in agreement with previous calculations~\cite{Cea2020Band,Long2022Atomistic,Long2023Electronic} and recent experimental results~\cite{Lai2023Imaging}. We emphasize that this is the \emph{only} twist angle in TBG for which one can have a perfect moir\'e pattern alignment (or a single moir\'e) with a hBN substrate. As this is only a geometrical condition, it is quite remarkable that it occurs practically at the magic angle where the bands in TBG tend to become flat.  
	
	In the particular case of pure biaxial strain, with no twist, Eq. \eqref{eq:Fbiaxial} reduces to $\mathbf{F}=\epsilon_{b}^{2}\mathbb{I}$. In fact, according to Eq. \eqref{eq:T} one simply has $\mathbf{g}_{i}=\epsilon_{b}\mathbf{b}_{i}$, i.e., the moiré vectors are just the reciprocal vectors scaled by the biaxial strain magnitude. Thus, in contrast to the cases of only a twist or shear strain, the moiré BZ for only biaxial strain has the same orientation as the BZ of the honeycomb lattices (see Fig. \ref{fig:momentumtransfer}). Similarly to the case of pure shear strain, the resulting moir\'e pattern has the same hexagonal periodicity as with only a twist angle $\theta_{eq}$ when
	\begin{equation}
		\epsilon_{b}=2\sin\left(\frac{\theta_{eq}}{2}\right).\label{eq:bitheta}
	\end{equation}
	However, the moir\'e orientation with only biaxial strain is rotated $90^{\circ}$ with respect to the case of only a twist angle, see Eq. \eqref{eq:alignB}. 
	
	A comparison between hexagonal moir\'e patterns formed by only a twist, and only shear or biaxial strain, can be seen in Fig. \ref{fig:TwistShearBiaxial}. Although in all situations the moir\'e patterns look practically the same at the moir\'e scale, the local distortions of each honeycomb lattice can be significantly different. Note that only in the cases of a pure twist or a pure biaxial strain, the moir\'e patterns have $\mathcal{C}_{3}$ rotational symmetry. 
	
	\subsection{Shear and biaxial strain}\label{subsec:ShearandBiaxial} 
	
	The combination of shear and biaxial strain implies that the lattices change both its size and shape. The general strain tensor of such combination,
	\begin{align}
		\mathcal{E} & =\left(\begin{array}{cc}
			\epsilon_{b} & 0\\
			0 & \epsilon_{b}
		\end{array}\right)+\mathrm{R}^{\mathrm{T}}\left(-\varphi\right)\left(\begin{array}{cc}
			0 & \epsilon_{s}\\
			\epsilon_{s} & 0
		\end{array}\right)\mathrm{R}\left(-\varphi\right)\nonumber \\
		& =\left(\begin{array}{cc}
			\epsilon_{b}-\epsilon_{s}\sin2\varphi & \epsilon_{s}\cos2\varphi\\
			\epsilon_{s}\cos2\varphi & \epsilon_{b}+\epsilon_{s}\sin2\varphi
		\end{array}\right),
	\end{align}
	leads to the transformation 
	\begin{align}
		\mathbf{F} & =4\sin^{2}\left(\frac{\theta}{2}\right)\mathbb{I}+\left(\epsilon_{s}^{2}+\epsilon_{b}^{2}\right)\cos^{2}\left(\frac{\theta}{2}\right)\mathbb{I}\nonumber \\
		& \;-2\epsilon_{s}\mathrm{R}\left(2\varphi\right)\left[\sin\left(\theta\right)\sigma_{z}-\epsilon_{b}\cos^{2}\left(\frac{\theta}{2}\right)\sigma_{x}\right],
		\label{eq:FSB}
	\end{align}
	The shear strain gives the non-spherical last term in $\mathbf{F}$, thus leading to non-hexagonal moiré patterns. The possible strained geometries are similar to those that result from uniaxial heterostrain (Section \ref{subsec:Uniaxial}). This analogy can be precisely stated by comparing the above transformation with Eq.~\eqref{eq:Funiaxial}. Indeed, since $\mathrm{R}\left(\phi\right)\sigma_{x}=\mathrm{R}\left(\phi+\pi/2\right)\sigma_{z}$, we have an equivalence of both transformations by the correspondence
	\begin{align}
		\epsilon & \rightarrow\epsilon_{s}+\epsilon_{b},\nonumber \\
		\nu & \rightarrow\frac{\epsilon_{s}-\epsilon_{b}}{\epsilon_{s}+\epsilon_{b}},\label{eq:SBcorr}\\
		\phi & \rightarrow\varphi+\pi/4.\nonumber 
	\end{align}
	This allows one to directly obtain the geometrical properties due to biaxial and shear strain from those previously studied for uniaxial heterostrain. For the particular case of equal length moiré vectors, the above correspondence can be replaced in the analytical expressions in Appendix~\ref{App: Equal}. We emphasize that the obtained moiré patterns, being either a result of uniaxial heterostrain, or a combination of shear and biaxial strain, are \emph{exactly} the same if the above correspondence holds. This can have important implications in the design of moiré patterns by strain, since it offers a wider range of strain configurations from which one can engineer them. Furthermore, it offers a more thorough description of the moiré patterns observed in experiments, since they do not necessarily could be the consequence of only uniaxial heterostrain.
	
	\begin{figure}[t]
		\includegraphics[scale=0.36]{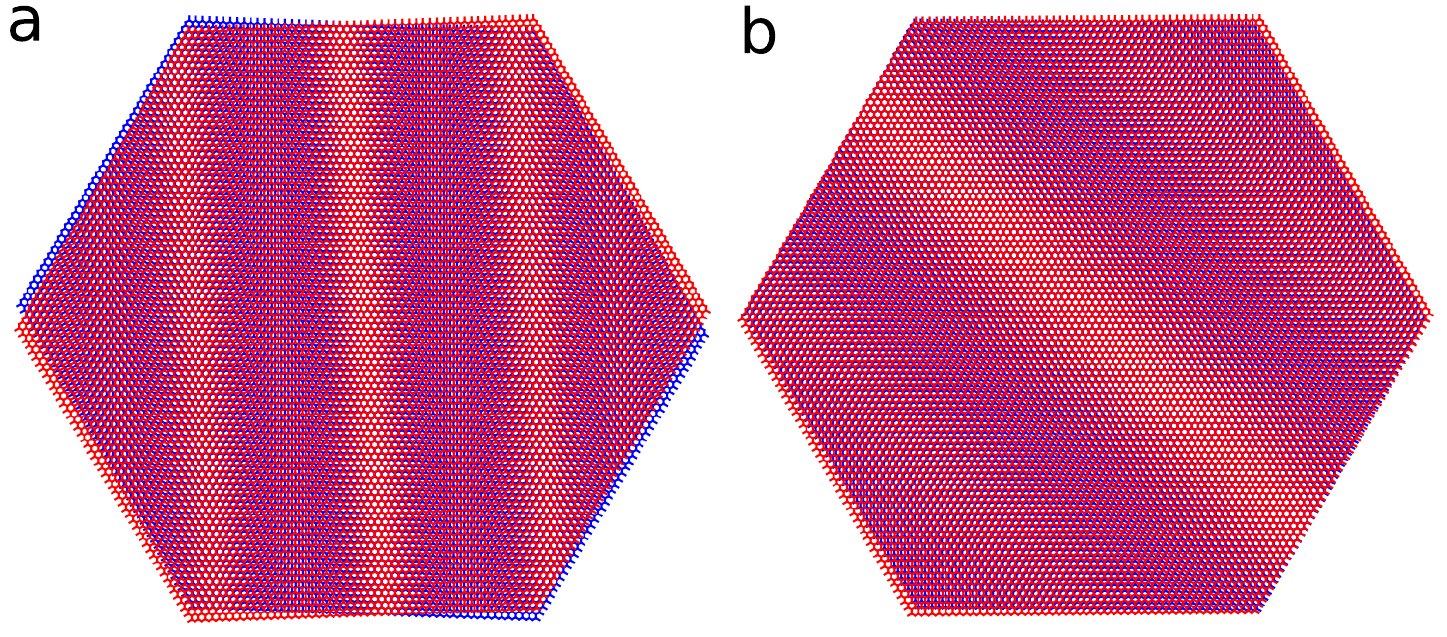}	
		\caption{Unidimensional channels formed by shear and biaxial strain, according to Eq. \eqref{eq:scriticalSB}. In (a) we consider a twist angle $\theta=2^{\circ}$, biaxial strain with magnitude $\epsilon_{b}=0.5\%$, and shear strain with magnitude $\epsilon_{s}\simeq\sqrt{\theta^{2}+\epsilon_{b}^{2}}\sim3.5\%$ and angle $\varphi=0^{\circ}$. In (b) there is no twist angle, and the channels arise solely from shear and biaxial strain with equal magnitude $\epsilon_{b}=\epsilon_{s}=1.5\%$.}\label{fig:1Dchannels}
	\end{figure}
	
	The correspondence given by Eq. \eqref{eq:SBcorr} directly extends to the critical limit in which the moiré vectors become collinear. Indeed, from Eq. \eqref{eq:scriticalU} we get that unidimensional channels arise under shear and biaxial strain if
	\begin{equation}
		\sqrt{\epsilon_{s}^{2}-\epsilon_{b}^{2}}=\pm2\tan\left(\frac{\theta}{2}\right),\label{eq:scriticalSB}
	\end{equation}
	independently of the shear angle $\varphi$. The same result is obtained from the condition $\mathrm{det}\,\mathbf{F}=0$ in Eq. \eqref{eq:FSB}. The above result extends the possibility of realizing quasi-undimensional channels in the presence of different strain configurations. Interestingly, these can result from only shear strain if
	\begin{equation}
		\epsilon_{s}=\pm2\tan\left(\frac{\theta}{2}\right).
	\end{equation}
	This critical shear strain is $\sim\sqrt{\nu}\simeq0.4$ times smaller than the one required for the case of uniaxial heterostrain. Eq. \eqref{eq:scriticalSB} further allows the possibility of unidimensional channels arising purely by strain, without a twist, which occurs when
	\begin{equation}
		\epsilon_{s}=\pm\epsilon_{b}.
	\end{equation}
	Remarkably, this condition does not depend on the strain magnitude: as long as they are non-zero, such combination would result in collinear moiré vectors. The shear angle $\varphi$ changes the orientation and length of the collinear moiré vectors (in real space, the orientation and stretch form of the channels). Two examples of unidimensional channels due to biaxial and shear strain are shown in Fig. \ref{fig:1Dchannels}, in the cases of with and without a twist angle.
	
	It is worth noting that the no-twist condition $\epsilon_{s}=\pm\epsilon_{b}$ relies on our initial assumption that the strain forces are equal but opposite in each honeycomb layer, i.e., $\mathcal{E_{\pm}}=\pm\mathcal{E}/2$ (Section \ref{subsec:GeneralStrain}). For arbitrary strain forces in each lattice, the transformation given by Eq. \eqref{eq:T} generalizes to $\mathbf{T}=\mathcal{E}_{+}-\mathcal{E}_{-}$ for the case of no twist. Then, for a combination of shear and biaxial
	strain (with, for simplicity, shear angle $\varphi=0$ in both layers), the undimensional channel condition $\mathrm{det}\,\mathbf{T}=0$ (see Ref. \cite{Sinner2023}) implies 
	\begin{equation}
		\epsilon_{s,+}-\epsilon_{s,-}=\pm\left(\epsilon_{b,+}-\epsilon_{b,-}\right)\label{eq:1Dgen}
	\end{equation}
	where $\epsilon_{s,\pm}$ and $\epsilon_{b,\pm}$ are the shear and biaxial strain magnitude in each layer. Thus there is, in general, quite a wide range of only-strain configurations that lead to unidimensional channels. It may be even possible that only one layer is strained, in which case Eq. \eqref{eq:1Dgen} is satisfied if $\epsilon_{s}=\pm\epsilon_{b}$. The generalization to arbitrary shear angles in each lattice further increases the possible only-strain configurations that yield unidimensional channels.
	
	\subsection{\textbf{Deformation of the Brillouin zone}}\label{sec:BZdeformation}
	
	In the reciprocal space, the most symmetrical primitive cell is given by the first Brillouin zone, which is constructed by considering the set of points that can be reached from the origin without crossing a Bragg plane (lines in the 2D case). The moir\'e patterns discussed in the previous sections imply that such cell would drastically change its shape under the application of strain. Consider, for example, the hexagonal BZ  of a honeycomb lattice. In terms of the reciprocal vectors, it can be obtained by the union of the points  
	\begin{align}
		\mathbf{q}_{1} & =-\frac{2\mathbf{g}_{1}+\mathbf{g}_{2}}{3},\nonumber\\
		\mathbf{q}_{2} & =\mathbf{q}_{1}+\mathbf{g}_{1},\label{eq:q}\\
		\mathbf{q}_{3} & =\mathbf{q}_{1}+\mathbf{g}_{1}+\mathbf{g}_{2},\nonumber
	\end{align}
	and their negatives. This construction holds in general for the moir\'e pattern of a twisted bilayer superlattice without strain, since then the two lattices are only relatively rotated. However, when the lattices are deformed, the construction through the six vectors $\mathbf{q}_{i}$ yields a deformed hexagon which is no longer the first BZ. The same holds for the moir\'e superlattice. 
	
\begin{figure}[t]
		\includegraphics[scale=0.13]{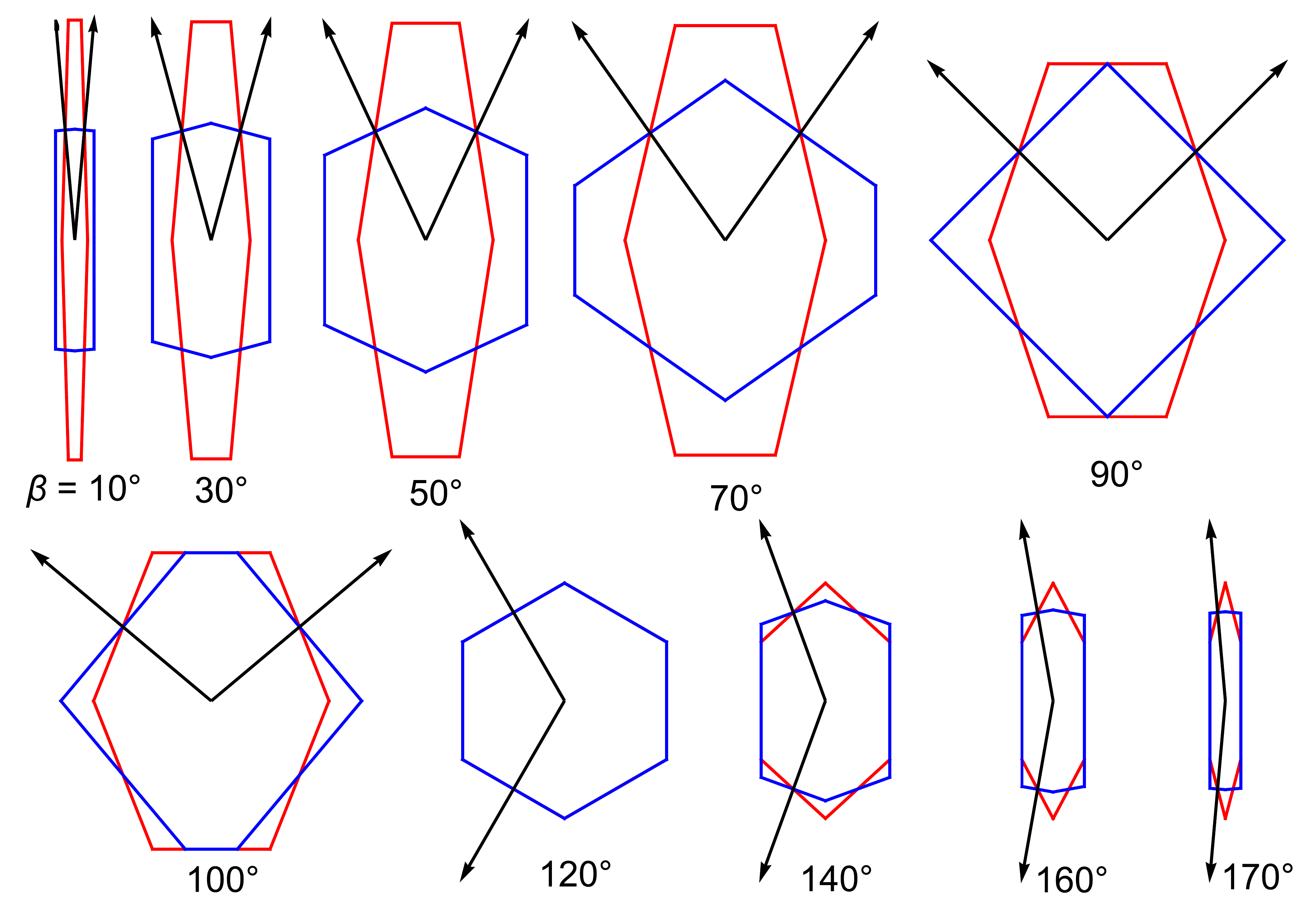}	
		\caption{Evolution of the Brillouin zone for different angles $\beta$ between equal length superlattice vectors (shown in black). Each respective mBZ constructed with vectors from Eqs. (\ref{eq:Q}) is shown in blue, while the deformed hexagons constructed using vectors from Eqs.~(\ref{eq:q}) is shown in red. Both constructions coincide only in the non-strain limit where $\beta=120{^\circ}$. With strain, the deformed hexagons do no longer capture the full symmetry of the moir\'e patterns. In particular, only the mBZ is symmetric around $\beta=90{^\circ}$, since it corresponds to the same moir\'e pattern rotated by $180{^\circ}$.}\label{fig:BZevol}
	\end{figure}

	Although the construction through Eq.~\eqref{eq:q} still gives a unit cell in reciprocal space, such cell does not reflect the symmetries of the strained moir\'e patterns. Specifically, we refer to the symmetries relating the AA and AB stacking positions of the moir\'e patterns, as seen in Fig.~\ref{fig:MoireFamily}. The correct construction of the moiré BZ (mBZ) under strain requires a generalization of Eq.~\eqref{eq:q} for the case in which the lattice vectors can have any angle and length. Following our previous discussion, we will focus on the situations in which the lattice vectors have equal length. In that case, the points that determine the mBZ are given by
	\begin{align}
		\mathbf{Q}_{1} & =-\frac{\left(1+2\chi\right)\mathbf{g}_{1}-\lambda\mathbf{g}_{2}}{2\left(1+\chi\right)},\nonumber\\
		\mathbf{Q}_{2} & =\mathbf{Q}_{1}+\mathbf{g}_{1},\label{eq:Q}\\
		\mathbf{Q}_{3} & =\mathbf{Q}_{1}+\mathbf{g}_{1}-\lambda\mathbf{g}_{2},\nonumber
	\end{align}
	where $\chi=\left|\mathbf{g}_{1}\cdot\mathbf{g}_{2}\right|/\left|\mathbf{g}_{1}\cdot\mathbf{g}_{1}\right|$ and $\lambda=\mathrm{sign}\left(\mathbf{g}_{1}\cdot\mathbf{g}_{2}\right)+\delta_{0,\mathbf{g}_{1}\cdot\mathbf{g}_{2}}$ (see the appendix~\ref{App: Const} for details).  It is easy to see that the points in Eq.~\eqref{eq:q} reduce to those given in the above equation only for a hexagonal lattice with $\beta=2\pi/3$. Note that for $\beta=\pi/2$ the six points are reduced to four because $\mathbf{Q}_{1}=-\mathbf{Q}_{3}$, thus resulting in a square mBZ. In Fig.~\ref{fig:BZevol} we show the evolution of the mBZ with the applied strain. A comparison is made with the deformed hexagon calculated with the points $\mathbf{q}_{i}$ given by Eq.~\eqref{eq:q}. 
	
	The mBZ, and its counterpart in real space (the Wigner--Seitz cell shown in Fig.~\ref{fig:MoireFamily}), provides a direct visualization of the geometrical properties of the moir\'e patterns under strain. This becomes clear by analyzing the shapes of the mBZ in Fig. \ref{fig:BZevol}, which follow a distinct pattern depending on the angle $\beta$. In contrast, the deformed hexagon cell only reflects the magnitude of the strain in the system (i.e., the larger the strain, the longer the deformed hexagon gets), similarly as how the AA stacking stretches in real space (see Fig. \ref{fig:MoireFamily}b). This behavior has been used to characterize the moir\'e patterns under strain, e.g., by reshaping the deformed hexagons to a regular form \cite{Fu2019}. We believe, however, that the alternative way of looking at the moir\'e patterns, by considering the mBZ or the Wigner--Seitz cell in real space, gives a clearer representation of the strained superlattice geometry. As noted in Sec.~\ref{subsec:GeneralStrain}, the underlying distortion of the honeycomb lattices, and thus of the magnitude of the strain, is reflected in the stretch of the AA stacking within the primitive cell. Furthermore, the reshaping of the mBZ may complement the approach in Ref.~\cite{Wang2023Open}, where strain-induced open Fermi surfaces in a distorted honeycomb cell were proposed to explain the unusual magnetotransport experiments in Ref.~\cite{Finney2022Unusual}. However, the impact of mBZ reshaping due to strain on magnetotransport experiments remains an open question. 
	
	\section{Electronic properties of strained moir\'e lattices} \label{sec:ElectronicProperties}
	
	\begin{figure}[t]
		\includegraphics[scale=0.075]{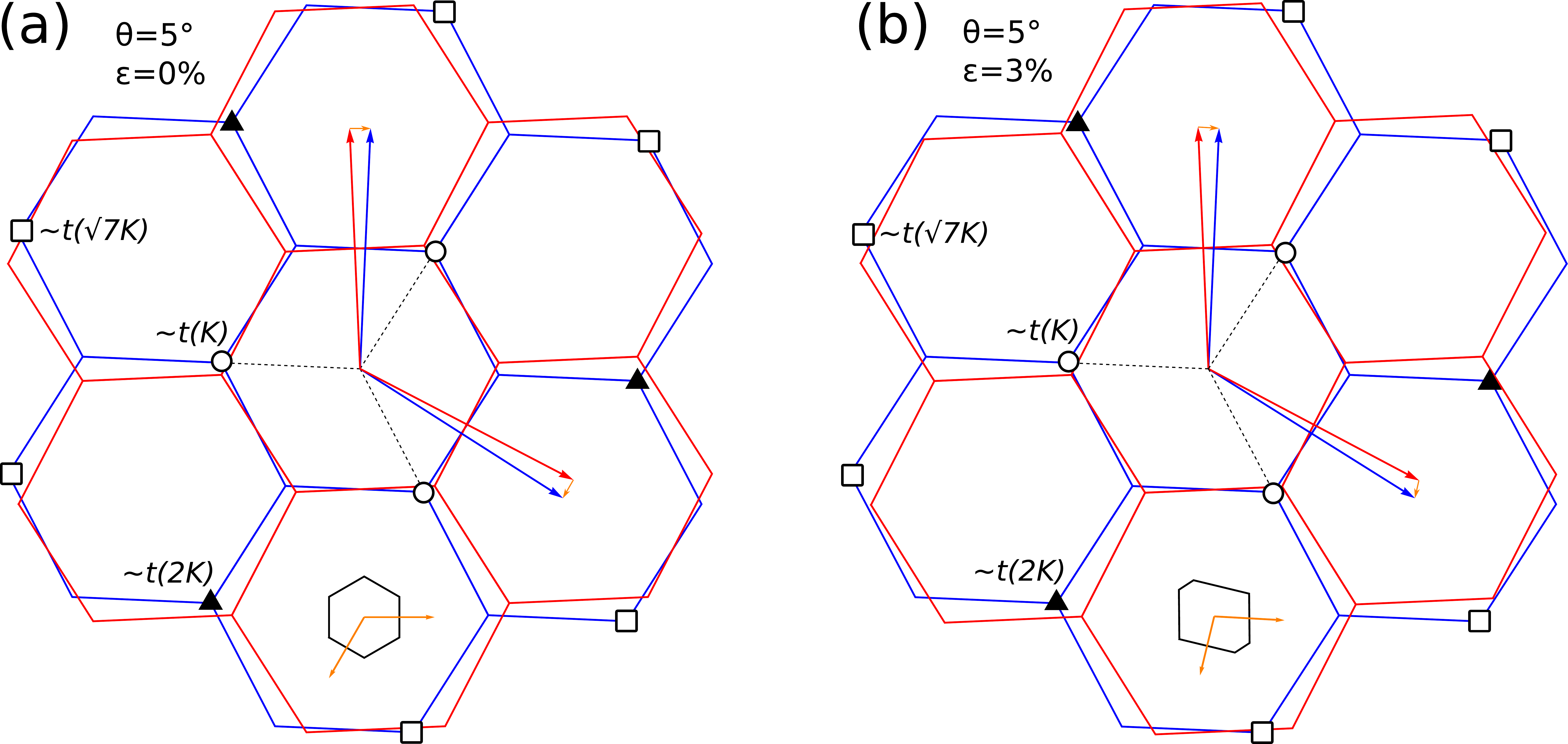}	
		\caption{Reciprocal space representation of moiré structures
			for the cases of (a) $\theta=5^{\circ}$, $\epsilon=0$ (no strain) and (b) $\theta=5^{\circ}$, with uniaxial heterostrain $\epsilon=3\%,\phi=0^{\circ}$. In both cases the three leading order Fourier contributions, for momenta relative to a $K$ point in the bottom layer (blue), are shown in open circles, filled triangles and open squares, with respective hopping magnitudes $\sim t\left(K\right)$, $\sim t\left(2K\right)$ and $\sim t\left(\sqrt{7}K\right)$. The corresponding moiré BZ in each case is shown in the bottom. Since the hopping magnitude $t\left(q\right)$ decays exponentially with
			$q$, for undeformed TBG one has $t\left(K\right)\gg t\left(2K\right)\gg t\left(\sqrt{7}K\right)$, which justifies keeping only the three Fourier components with magnitude $\sim t\left(K\right)$. This still holds under small strain, as each lattice is only slightly distorted. Such small deformations can, nevertheless, significantly reshape the moiré geometry and BZ.}\label{fig:MoireC}
	\end{figure}
	
	\begin{figure*}[t]
		\includegraphics[width = \textwidth]{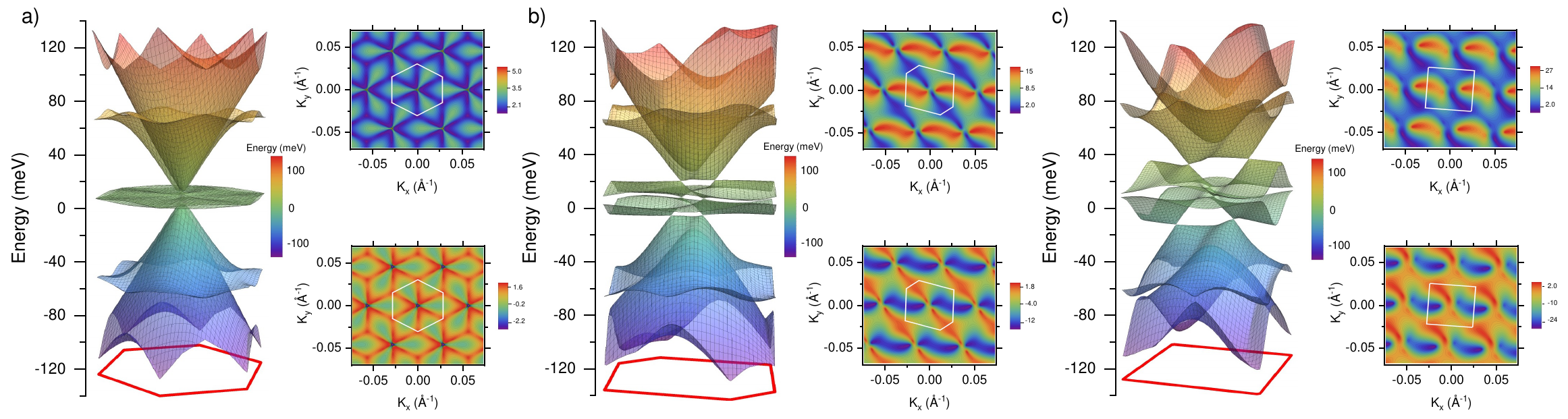}	
		\caption{Evolution of the band structure of heteroaxially strained TBG with a twist angle $\theta=1.05^{\circ}$, for different angles $\beta$ between equal length moir\'e vectors. The strain parameters are: a) $\epsilon=0$ (no strain), b) $\epsilon\simeq 0.4\%$, $\phi=-12.0^{\circ}$ and c) $\epsilon\simeq 0.8\%$, $\phi=-9.40^{\circ}$. The angles between the reciprocal moiré vectors are $\beta = 120^{\circ}, 105^{\circ} \text{ and } 90^{\circ}$, respectively. Underneath each 3D plot we show the corresponding mBZ, and to their side the corresponding density plot of the lower (bottom panel) and upper (top panel) middle band. The respective mBZ constructed with vectors defined in Eq.~(\ref{eq:Q}) are shown in white.}
		\label{fig:BandStructure}
		\label{fig:3dbands}
	\end{figure*}
	
	\subsection{Effective continuum models}
	
	While the shape and form of strained moir\'e patterns only reflect the geometrical differences between deformed lattices, the electronic properties reflect other important consequences of the strain, such as the shift of the Dirac points, the influence of the moir\'e potential that couples them, and the splitting of the van Hove singularities, among others~\cite{Metal21,Huder2018Hetero}. For sufficiently low twist angles, these properties can be captured by a direct extension of the continuum model~\cite{LopesdosSantos2007, Bistritzer2011, LopesdosSantos2012} in the presence of strain~\cite{Metal21,Huder2018Hetero,Fu2019}.
	
	First, we note that under strain the mBZ and the position of the Dirac points in each lattice change. As a result, the latter in general do not coincide with the high symmetry points at borders of the mBZ. At small deformations, the new positions of the Dirac points in the $\zeta$ valley of the $\ell=\pm$ layer are given by
	\begin{equation}
		\mathbf{D}_{\ell,\zeta}\simeq\left(\mathbb{I}-\ell\mathcal{E}/2\right)\mathrm{R}\left[\ell\theta/2\right]\mathbf{K}_{\zeta}-\ell\zeta\mathbf{A},\label{eq:DPshift}
	\end{equation}
	where $\mathbf{K}_{\zeta}=-\zeta\left(2\mathbf{b}_{1}+\mathbf{b}_{2}\right)/3$ is the Dirac point in the undeformed honeycomb lattice, and 
	\begin{equation}
		\mathbf{A}=\frac{\sqrt{3}}{2a}\beta\left(\epsilon_{xx}-\epsilon_{yy},-2\epsilon_{xy}\right)
	\end{equation} 
	is the strain-induced gauge potential ($\beta_{G}\simeq3.14$ is the Grüneisen parameter)~\cite{VKG10}. The two terms in Eq. \eqref{eq:DPshift} represent the combined effect of the strain on the position of the Dirac points: the first term gives the shift due to the geometrical distortion of the lattice, while the second term gives the shift due to the change in the hopping energies. In addition, strains can also lead to scalar (or deformation) potentials~\cite{Manes2007Symmetry, Suzuura2002Phonons,VKG2010}
	\begin{equation}
		V=g\left(\epsilon_{xx}+\epsilon_{yy}\right).
		\label{eq: DeformationPote}
	\end{equation} 
	We use $g=4$ eV for monolayer graphene~\cite{Choi2010Effects}. The aforementioned potential is incorporated into the diagonal elements of the Dirac Hamiltonian, resulting in a vertical energy displacement of the Dirac cones within each monolayer. This phenomenon resembles the responses observed under the influence of a perpendicular electric field~\cite{Moon2014Optical}. 
	
	In a TBG configuration, at low twist angles and strain magnitudes the low energy physics is dominated by states near the shifted Dirac points $\mathbf{D}_{\ell,\zeta}$. The continuum model Hamiltonian under strain, for the $\zeta$ valley, then reads
	\begin{equation}
		H_{\zeta}=\left[\begin{array}{cc}
			h_{-,\zeta}\left(\mathbf{q}\right)-\mathbb{I}V & U^{\dagger}\left(\mathbf{r}\right)\\
			U\left(\mathbf{r}\right) & h_{+,\zeta}\left(\mathbf{q}\right)+\mathbb{I}V
		\end{array}\right].
	\end{equation}
	Here $h_{\ell,\zeta}\left(\mathbf{q}\right)$ is the Dirac Hamiltonian in the $\ell$ layer,
	\begin{equation}
		h_{\ell,\zeta}\left(\mathbf{q}\right)=-\hbar v_F\boldsymbol{\sigma}_{\zeta}\cdot\mathrm{R}^{\mathrm{T}}\left(\ell\theta/2\right)\left(1+\ell\mathcal{E}/2\right)\left(\mathbf{q}-\mathbf{D}_{\ell,\zeta}\right),
	\end{equation}
	where $v_F$ is the Fermi velocity and $\boldsymbol{\sigma}_{\zeta}=\left(\zeta\sigma_{x},\sigma_{y}\right)$. 
	
	\begin{figure*}[t]
		\includegraphics[scale=0.18]{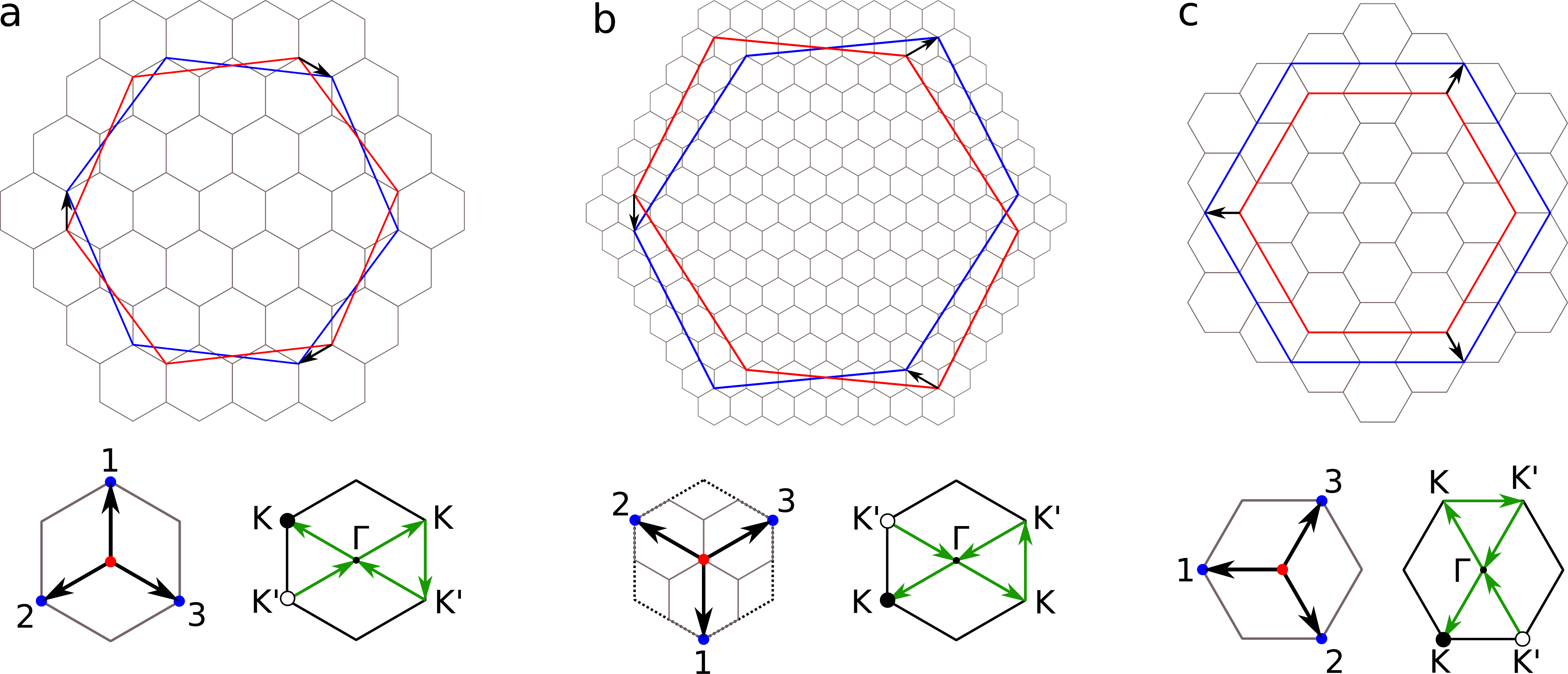}	
		\caption{Reciprocal space representation of superlattice structures with hexagonal unit cells. Large hexagons represent the BZ of each monolayer (in red and blue). The moiré BZ are represented by the small black hexagons. The figure shows the structures for: a) pure twist angle, b) pure shear strain and c) pure biaxial strain. The corresponding mBZ and the hopping processes between the Dirac cones of each graphene monolayer are displayed at the bottom. Arrows indicate the direction of the momentum transfers between Dirac points.}
		\label{fig:momentumtransfer}
	\end{figure*}
	
	The coupling between the layers is given by the matrix $U\left(\mathbf{r}\right)$
	in the non-diagonal terms of $H_{\zeta}$. For long-period moiré structures
	its Fourier expansion reads \cite{Koshino2015}
	\begin{align}
		U_{\alpha\beta}\left(\mathbf{r}\right) & =\sum_{m_{1},m_{2}}e^{i\left(m_{1}\mathbf{g}_{1}+m_{2}\mathbf{g}_{2}\right)\cdot\mathbf{r}}e^{i\left(m_{1}\mathbf{b}_{1}+m_{2}\mathbf{b}_{2}\right)\cdot\left(\boldsymbol{\delta}_{\alpha}-\boldsymbol{\delta}_{\beta}\right)}\nonumber \\
		& \qquad\times t_{\alpha\beta}\left(\mathbf{k}_{-}+m_{1}\mathbf{b}_{1,-}+m_{2}\mathbf{b}_{2,-}\right).\label{eq:Ufourier}
	\end{align}
	Here the indices $\alpha$ and $\beta$ refer to the sublattices $A$ and $B$ in each layer, with $\boldsymbol{\delta}_{\alpha,\beta}$ being the corresponding basis vectors. The interaction strength is determined by the hopping parameter $t_{\alpha\beta}\left(\mathbf{k}_{-}+m_{1}\mathbf{b}_{1,-}+m_{2}\mathbf{b}_{2,-}\right)$ which, importantly, only depends on the strained reciprocal vectors in one layer. For momenta relative to a $K$ point, the coupling amplitudes thus scale as $\sim t\left(K\right)$ to first order, $\sim t\left(2K\right)$ to second order, and so on (see Fig. \ref{fig:MoireC}). For TBG it was estimated that \cite{Koshino2015} $t\left(K\right)\sim110$ meV and $t\left(2K\right)\sim1.6$ meV, which justifies keeping in Eq. \eqref{eq:Ufourier} only the three leading order terms with amplitude $t\left(K\right)$. Assuming, on the basis of small deformations, that under strain one still has $t\left(2K\right)/t\left(K\right)\ll1$, the leading order Fourier expansion of the moiré coupling matrix around the $K$ point at $\mathbf{K}_{-}=-\zeta\left(2\mathbf{b}_{1,-}+\mathbf{b}_{2,-}\right)/3$ then reads \cite{Fu2019}
	\begin{equation}
		U\left(\mathbf{r}\right)\simeq U_{0}+U_{1}e^{i\zeta\mathbf{g}_{1}\cdot\mathbf{r}}+U_{1}^{\mathrm{T}}e^{i\zeta\left(\mathbf{g}_{1}+\mathbf{g}_{2}\right)\cdot\mathbf{r}},
	\end{equation}
	where
	\begin{equation}
		U_{0}=\left(\begin{array}{cc}
			u_{0} & u_{1}\\
			u_{1} & u_{0}
		\end{array}\right),\quad U_{1}=\left(\begin{array}{cc}
			u_{0} & u_{1}\omega^{-\xi\zeta}\\
			u_{1}\omega^{\xi\zeta} & u_{0}
		\end{array}\right).
	\end{equation}
	Here $\omega=e^{i2\pi/3}$, and $u_{1}$, $u_{0}$ are the $AB$ and $AA$ hopping energies, respectively. For the numerical calculations, we use $u_{1} = u_{2} = 90$ meV and $\hbar v_{F}/a = 2.135$ eV. In the matrix $U_{1}$ we have introduced a factor $\xi=\pm1$ that accounts for the phase of the three leading order momentum transfers between the shifted Dirac points in each layer. This phase is contingent upon the specific type of strain. In particular,  $\xi=1$ in Fig.~\ref{fig:momentumtransfer}a) and c) for TBG and pure biaxial strain, respectively, and $\xi=-1$ in Fig.~\ref{fig:momentumtransfer}b) for pure shear strain. Besides this phase factor, and a possible rescaling of the hopping energies, the coupling matrices $U_0$ and $U_1$ have the same form as in TBG \cite{Moon2014Optical}. The strain influence on the moiré coupling comes mainly from the modification of the moiré vectors $\mathbf{g}_i$ in its Fourier expansion, and thus of the momentum transfer vectors $\mathbf{q}_i$ between the Dirac points in each layer.
	
	\begin{figure*}[t]
		\includegraphics[width = \textwidth]{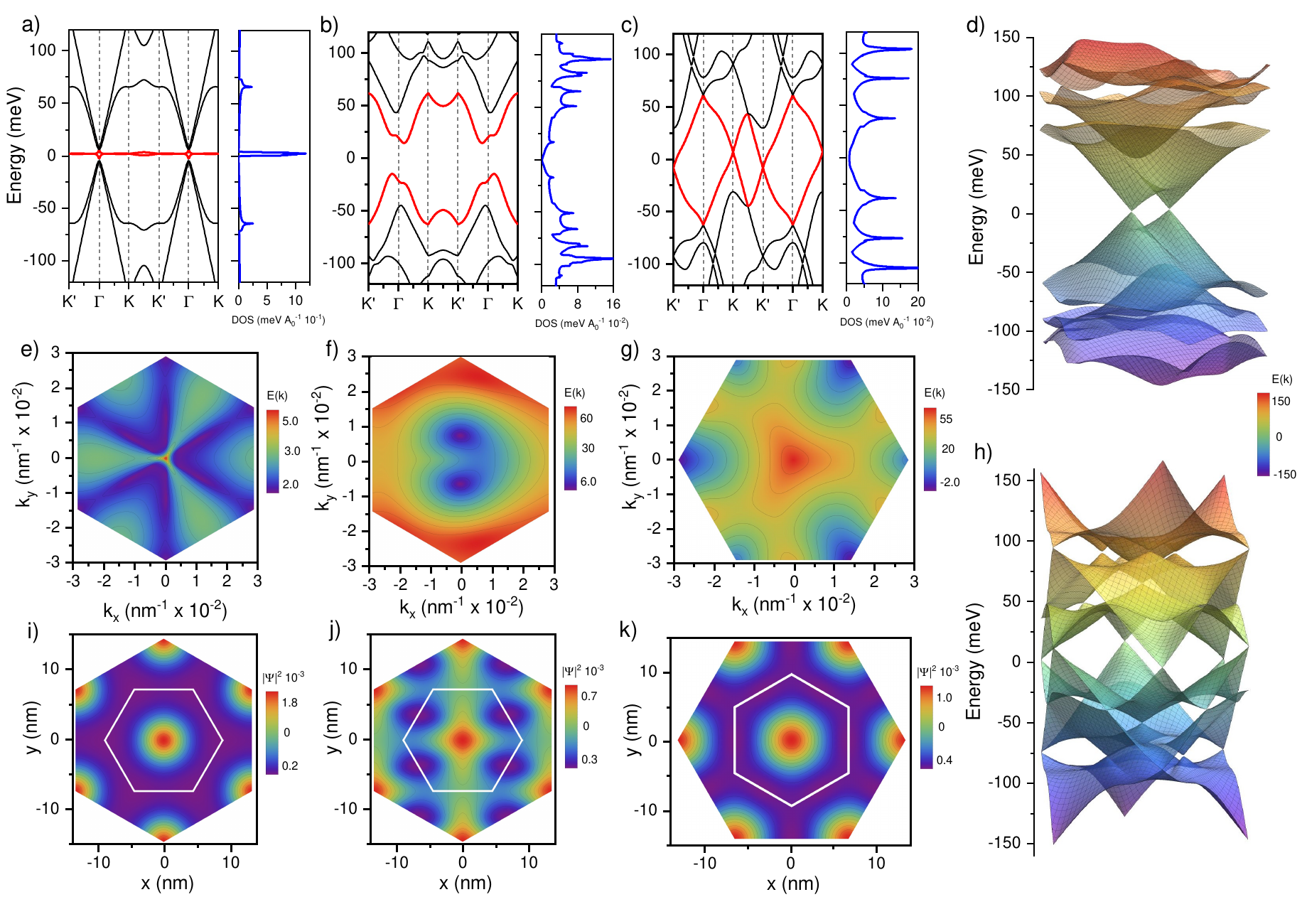}	
		\caption{Band structures of hexagonal moir\'e patterns generated by: (a) only twist angle $\theta=1.05^{\circ}$; (b) only shear strain with the magnitude $\epsilon_{s}=2\sin\left(\theta/2\right)\simeq1.83\%$; (c) only biaxial strain with the magnitude $\epsilon_{b}=2\sin\left(\theta/2\right)\simeq1.83\%$. The momentum path in each case is shown in Fig. \ref{fig:momentumtransfer}. From Eqs. \eqref{eq:exytheta} and \eqref{eq:bitheta}, all case have the same moir\'e periodicity $L \simeq13.4$ nm. Panels e)-g) display the corresponding density plot for the top middle band, while panels i)-k) display the total charge density profile of the lower middle band. 3D plots in panels d) and h) show the bands for the moir\'e structures realized by the shear and biaxial strain, respectively.  In the case of biaxial strain the mBZ and the Winger--Seitz cell are rotated $90^{\circ}$ degrees with respect to the other cases (cf. Fig. \ref{fig:momentumtransfer}c) and have the same orientation as a monolayer graphene on hBN.}
		\label{fig:HexaBands}
	\end{figure*}
	
	\subsection{Electronic structure: twist and strain}
	
	We first consider the case of TBG with uniaxial heterostrain~\cite{Huder2018Hetero,Fu2019}. The numerical results for the band structure are shown in Fig.~\ref{fig:BandStructure}.  
	As can be seen, even for relative low strain magnitudes the band structure can greatly differ from the one in the non-strain case. The discussion of several aspects is in order. First, we note that under strain the positions of the shifted Dirac points define a periodicity which does not coincide anymore with the corners of the mBZ. Indeed, according to Eq. \eqref{eq:DPshift}, the difference $\Delta \mathbf{D}=\mathbf{D}_{-}-\mathbf{D}_{+}$ between two shifted Dirac points corresponding to, e.g., the non-deformed position $\mathbf{K}=-\left(2\mathbf{b}_{1}+\mathbf{b}_{2}\right)/3$, is given by
	\begin{equation}
		\Delta\mathbf{D}=-\frac{2\mathbf{g}_{1}+\mathbf{g}_{2}}{3}+2\mathbf{A},
	\end{equation}
	where $\mathbf{g}_{i}=\mathbf{T}\mathbf{b}_{i}$ are the strained moir\'e vectors. Clearly, the vector $\Delta\mathbf{D}$ only coincides with the corner $\mathbf{Q}_{1}$ of the mBZ [cf. Eq.~\eqref{eq:Q}] when the angle between $\mathbf{g}_{1}$ and $\mathbf{g}_{2}$ is $120^{\circ}$ and $\mathbf{A}=0$, i.e., the non-strain case. Note that even in the case of pure shear strain with no twist, where the moir\'e geometry is the same as in the only-twist case, the vector $\Delta\mathbf{D}$ would still be shifted from the hexagonal mBZ due to the non-zero gauge field $\mathbf{A}_{\mathrm{shear}}\propto\left(0,-2\epsilon_{xy}\right)$. This is expected because the honeycomb lattices are distorted due to the strain, and therefore the hopping energies are no longer the same as in the only twist case. It should be also noted that any relation between the Dirac points and the borders of the moir\'e BZ is further blurred at low twist angles, where the Dirac points are strongly coupled by the moir\'e potential.
	
	Besides the actual shift in momentum due to strain induced gauge and deformation fields, there is also an additional energy shift of the Dirac points, which gets larger as the strain increases. As a result, the lowest bands around the magic angle still have two distinct Dirac points in the presence of strain. A close inspection reveals that such suppression of the flat bands occurs even when the gauge and deformation fields are not taken into account (cf. Appendix \ref{sec:NoGauge}), thus hinting that it is mainly due to how the strain influences the coupling of the Dirac points by the moiré potential. We have observed such flat suppression not only at the magic angle $\theta\sim 1.05^{\circ}$ with no strain (as shown in Fig. \ref{fig:BandStructure}), but also for \textit{any} other combinations of twist and strain. In other words, our results seem to indicate that the strain \textit{does not} shift the magic angle to a new value, or leads to new flat bands conditions compared to unstrained TBG. Although a concrete explanation of this behavior is still lacking, it may hint that the origin of flat bands in TBG is intrinsically related to the symmetries of the system, particularly those relating the moiré potential $U\left(\mathbf{r}\right)$ (which always has a hexagonal symmetry), and the three momentum transfers $\mathbf{q}_i$ (whose hexagonal symmetry is in general broken by the strain). Note that, although the strain breaks $\mathcal{C}_{3z}$, $\mathcal{C}_{2x}$ and $\mathcal{C}_{2y}$ rotational symmetries, the symmetry $\ensuremath{\mathcal{C}_{2z}\mathcal{T}}$, with $\mathcal{T}$ a time reversal operator, remains intact~\cite{Metal21,Fu2019}, so that the Dirac cones are not gapped by strain,  as seen in Fig. \ref{fig:BandStructure} and Fig.~\ref{fig:HexaBands}.
	
	\subsection{Electronic structure: Pure strain}
	
	Next we examine the scenario of hexagonal moir\'e structures emerging solely from strain (cf. Fig.~\ref{fig:TwistShearBiaxial}). These cases are interesting because, when compared to the situation of hexagonal patterns arising from only a twist, they reflect the direct effect of strain in the electronic properties. In particular, by using the relations~\eqref{eq:exytheta} and~\eqref{eq:bitheta}, we are able to compare the electronic structures of cases that share the same moir\'e periodicity. In Fig.~\ref{fig:HexaBands} we present the results for the band structure, density of states and charge density. For comparison, we also include the results for TBG without strain.
	
	Remarkably, although all cases shown have the same hexagonal moir\'e periodicity, their electronic properties differ substantially. The strain thus plays a decisive role in how the Dirac points in each lattice couple through the moir\'e potential. This can be attributed to the actual distortion of each lattice under strain, which, as seen in Fig.~\ref{fig:TwistShearBiaxial}, results in different behaviors around AA or AB stacking positions, even if at the moir\'e scale they all look the same. Within the continuum model, these differences are mainly reflected in how are the three leading hopping processes between the Dirac points in each lattice, cf. Fig.~\ref{fig:momentumtransfer}. In particular, we only observe flat bands, and a corresponding peak in the density of states, in unstrained TBG. With strain, these flat bands disappear, and a splitting and emergence of multiple high-order van Hove singularities takes place. The overall influence of the strain can be more clearly seen in the density plot of the band structures. Note that in the case of only shear strain the two Dirac points are shifted and no longer captured along the momentum path depicted in Fig. \ref{fig:momentumtransfer}(b). 
	
	As in the case of twist and strain, there does not seem to be a new flat-band condition for the only-strain cases. Indeed, we have not found an equivalent twist angle where the bands flatten as in unstrained TBG. This may further hint that flat band realization is actually related to the orientation of the moiré vectors $\mathbf{g}_i$ with respect to the fixed angles in the moiré coupling matrices $U_i$, since in all the three cases shown in Fig. \ref{fig:HexaBands} the momentum transfer vectors only differ in their orientation (cf. Fig. \ref{fig:momentumtransfer}). It should be noted that the observed behavior is restricted to moiré structures arising from graphene-like honeycomb layers, with a Dirac dispersion. 
	In other superlattices configurations, as e.g. in strain-only transition metal dichalcogenide moiré homobilayers, the strain may facilitate the formation of flat bands~\cite{Fu2019,Kundu2021}. 
	
	In Fig. \ref{fig:HexaBands} we also observe that in TBG the difference between the charge density at the center and at the edges of the mBZ is more significant than in the two cases involving only strain. This contrast implies potential variations in the electrostatic interactions within purely strained systems when compared to those observed in TBG~\cite{Guinea2018Electrostatic,Cea2022Electrostatic}. We note that our continuum model results for biaxial strain are in agreement with recent DFT calculations in strained untwisted graphene bilayers~\cite{Liu2023Exotic}, where the shift of the Dirac cones with strain indicates the presence of scalar deformation potentials, c.f. Eq.~\eqref{eq: DeformationPote}.   
	
	\section{Conclusions}\label{sec:conclusions}
	
	We have presented a general theoretical scheme that describes the strain effects in twisted two-dimensional materials. We have shown that the interplay between twist and strain can lead to the formation of practically any moiré geometry. The strain plays a central role in this by distorting the lattices and thus modifying the resulting relative length and angle between the moiré vectors. Due to the magnifying effect of the moiré pattern formation, this effect becomes significant even at very small strain magnitudes, where each layer’s lattice is barely deformed. Thus the plethora of moiré patterns observed in experiments can be directly attributed to the presence of small strain in the samples. Our considerations, however, go far beyond the mere diagnosis of such intrinsic effects and offer a platform to actually design moiré patterns by strain. Indeed, we have described in details the necessary conditions to form any desired moiré geometry, simply by selectively changing the twist and strain parameters. In particular, we have specified the conditions to form special moiré geometries, such as square moiré patterns, or hexagonal moiré patterns induced solely by strain. Furthermore, we have identified that the modifications of the moiré geometry due to the strain lead to significant deformations of the moiré Brillouin zone (mBZ). In contrast to previous studies we have found that, when subject to strain, the mBZ is not a deformed stretched hexagon, but rather a primitive cell that reflects new symmetries of the strained moiré vectors. This might have important implications, in particular with respect to identifying the high symmetry points in band structures. We have rounded up our studies by analyzing the electronic properties of the above strained moiré pattern. We have found that the strain seems to suppress the formation of moiré flat bands, even in those hexagonal patterns formed only by strain. It also tends to split and induce higher order van Hove singularities, as well as to modify the charge density profile. 
	
	\begin{acknowledgments}
		
		We thank Vincent Renard, Niels R. Walet, Adrian Ceferino and Gerardo Naumis for discussions. IMDEA Nanociencia acknowledges support from the \textquotedblleft Severo Ochoa\textquotedblright ~Programme for Centres of Excellence in R\&D (Grant No. SEV-2016-0686). F.E. acknowledges support from a research fellowship of CONICET (Argentina
		National Research Council). Z.Z. acknowledges support funding from the European Union's Horizon 2020 research and innovation programme under the Marie Skłodowska-Curie grant agreement No 101034431 and from the ``Severo Ochoa" Programme for Centres of Excellence in R\&D (CEX2020-001039-S / AEI / 10.13039/501100011033). P.A.P and F.G. acknowledge funding from the European Commission, within the Graphene Flagship, Core 3, grant number 881603 and from grants NMAT2D (Comunidad de Madrid, Spain), SprQuMat and (MAD2D-CM)-MRR MATERIALES AVANZADOS-IMDEA-NC.
		
	\end{acknowledgments}
	
	\appendix
	
	\section{Hexagonal symmetry in strained moir\'e patterns}\label{App: HexagonalS}
	As discussed in Sec. \ref{subsec:GeneralStrain}, and shown in an example in Fig. \ref{fig:mvectors}, under strain the construction of the moir\'e vectors using the usual recipe $\mathbf{g}_{i}=\tilde{\mathbf{b}}_{-,i}-\tilde{\mathbf{b}}_{+,i}$ does not always yield the smallest (primitive) vectors of the superlattice. A comprehensive analysis of all moir\'e geometries that can be formed under strain should account for which construction of the moir\'e vectors give the smallest one. The analysis can be simplified by using symmetry arguments. 
	
	First we note that the smallest primitive moir\'e vectors are, in general, given by one of these vectors: $\mathbf{g}_{1}$, $\mathbf{g}_{2}$, or $\left(\mathbf{g}_{1}+\mathbf{g}_{2}\right)$, where $\mathbf{g}_{i}=\mathbf{T}\mathbf{b}_{i}$ [cf. Eq. \eqref{eq:T}]. Choosing the first two gives, of course, the usual set $\left\{ \mathbf{g}_{1},\mathbf{g}_{2}\right\} $, from which one can carry out the analysis of the strained moir\'e geometries as done in the manuscript. But there are also two other possible sets: $\left\{ \mathbf{g}_{1},\mathbf{g}_{1}+\mathbf{g}_{2}\right\} $ and $\left\{ \mathbf{g}_{2},\mathbf{g}_{1}+\mathbf{g}_{2}\right\} $ (see Fig. \ref{fig:mvectors}). One of these two sets can be primitive when a translation of one moir\'e vector $\mathbf{g}_{i}$ by the other one $\mathbf{g}_{j}$ results in a smaller vector (i.e., when $\left|\mathbf{g}_{1}+\mathbf{g}_{2}\right|<\left|\mathbf{g}_{i}\right|$). The appropriate construction of the primitive moir\'e vectors by one of these three sets preserves the symmetries of the system (in the present case, the hexagonal symmetry of the underlying honeycomb lattices).
	
	\begin{figure}[t]
		\includegraphics[scale=0.25]{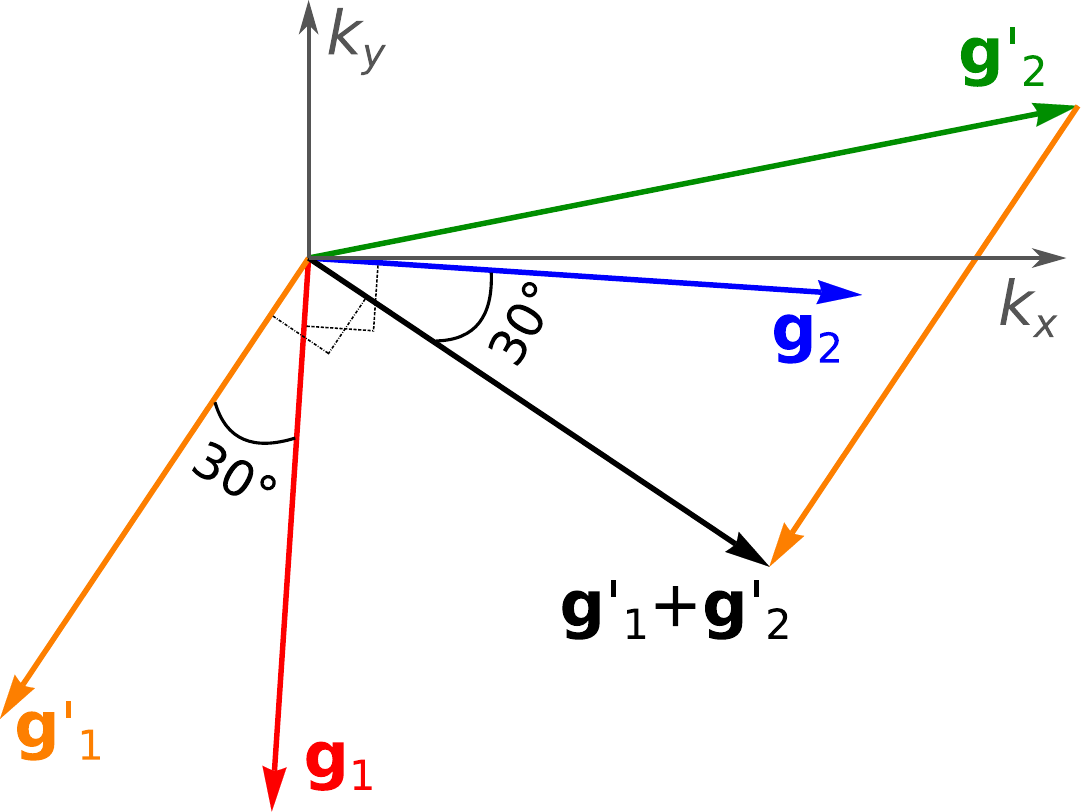}	
		\caption{Symmetries of the square moir\'e pattern solutions for $\theta=2^{\circ}$, as given by Eq. \eqref{eq:Eeq}. The depicted cases correspond to $\epsilon\simeq0.94\tan\left(\theta/2\right)$, with $\phi\simeq-9.4^{\circ}$ (reciprocal moir\'e vectors $\mathbf{g}_{1}$ and $\mathbf{g}_{2}$), and $\phi'=\phi+60^{\circ}$ (reciprocal vectors $\mathbf{g}'_{1}$ and $\mathbf{g}'_{2}$). First case leads straightforwardly to the perpendicular moir\'e vectors (cf. Fig. \ref{fig:MoireFamily}a). However, in the latter case the vectors $\mathbf{g}'_{1}$ and $\mathbf{g}'_{2}$ are not primitive, and the angle between them is not $90^{\circ}$. The resulting superlattice is still a square, as seen by the translation of $\mathbf{g}'_{2}$ by $\mathbf{g}'_{1}$ that yields the vector $\left(\mathbf{g}'_{2}+\mathbf{g}'_{1}\right)$ perpendicular to $\mathbf{g}'_{1}$. Therefore, the symmetric case with $\phi'=\phi+60^{\circ}$ also gives, as expected, a square moir\'e pattern, except that the respective primitive moir\'e vectors are to be constructed from the difference between the strained reciprocal vectors $\mathbf{b}_{1}$ and $\mathbf{b}_{1}+\mathbf{b}_{2}$ in each lattice.}\label{fig:Msquare}
	\end{figure}
	
	To see this, consider the equal length moir\'e vectors condition for the case of uniaxial heterostrain. For the usual set $\left\{ \mathbf{g}_{1},\mathbf{g}_{2}\right\}$, such condition implies the strain magnitude given by Eq. \eqref{eq:Eeq}. For the other two sets $\left\{ \mathbf{g}_{1},\mathbf{g}_{1}+\mathbf{g}_{2}\right\} $ and $\left\{ \mathbf{g}_{2},\mathbf{g}_{1}+\mathbf{g}_{2}\right\} $, the equal length condition can be stated as
	\begin{align}
		\mathbf{F}\left[\mathbf{b}_{1}-\left(\mathbf{b}_{1}+\mathbf{b}_{2}\right)\right]\cdot\left[\mathbf{b}_{1}+\left(\mathbf{b}_{1}+\mathbf{b}_{2}\right)\right] & =0,\\
		\mathbf{F}\left[\mathbf{b}_{2}-\left(\mathbf{b}_{1}+\mathbf{b}_{2}\right)\right]\cdot\left[\mathbf{b}_{2}+\left(\mathbf{b}_{1}+\mathbf{b}_{2}\right)\right] & =0,
	\end{align}
	where we have used that $\mathbf{F}=\mathbf{T}^{\mathrm{T}}\mathbf{T}$ is a symmetric transformation. Solving for the strain magnitude gives
	\begin{align}
		\left|\mathbf{g}_{1}\right|=\left|\mathbf{g}_{1}+\mathbf{g}_{2}\right| & \rightarrow\epsilon_{\mathrm{eq,2}}=\frac{4}{\nu-1}\cot\left(2\phi\right)\tan\left(\theta/2\right),\\
		\left|\mathbf{g}_{2}\right|=\left|\mathbf{g}_{1}+\mathbf{g}_{2}\right| & \rightarrow\epsilon_{\mathrm{eq,3}}=\frac{4}{\nu-1}\cot\left(\frac{\pi}{3}+2\phi\right)\tan\left(\theta/2\right).
	\end{align}
	Comparing with Eq. \eqref{eq:Eeq} we see that
	\begin{align}
		\epsilon_{\mathrm{eq}}\left(\phi+\pi/3\right) & =\epsilon_{\mathrm{eq,3}}\left(\phi\right),\\
		\epsilon_{\mathrm{eq}}\left(\phi-\pi/3\right) & =\epsilon_{\mathrm{eq,2}}\left(\phi\right),
	\end{align}
	thus restoring the hexagonal symmetry. Geometrically, the obtained result means that for given parameters $\left(\theta, \epsilon_{\mathrm{eq}},\phi\right)$, one of the equal length moir\'e vector is no longer primitive after the transformation $\phi\rightarrow\phi\pm\pi/3$. Rather, the new primitive vector is found by appropriately changing the moir\'e vector construction (e.g., by using other set than the usual one $\left\{ \mathbf{g}_{1},\mathbf{g}_{2}\right\} $). We emphasize that this is only a change in the superlattice description, due to how the moir\'e vectors are constructed. The observed moir\'e geometry, arising from the superposition of two strained honeycomb lattices with primitive vectors $\tilde{\mathbf{a}}_{i,\pm}=\left(\mathbb{I}+\mathcal{E}_{\pm}\right)\mathrm{R}\left(\pm\theta/2\right)\mathbf{a}_{i}$, always reflects the honeycomb symmetries of the underlying lattices, such that any translation $\phi\rightarrow\phi\pm\pi/3$ leads to the same moir\'e pattern (up to an overall rotation of the system). For this reason it is more convenient to study, as done in the manuscript, the strained moir\'e patterns by using only the set of vectors $\left\{ \mathbf{g}_{1},\mathbf{g}_{2}\right\} $, and generalizing the obtained results by taking into account the missing solutions corresponding to translations $\phi\rightarrow\phi+\pi/3$. These latter solutions would then correspond to the ones obtained by considering the other sets of possible primitive moir\'e vectors.
	
	\section{Analytical solutions for equal length moir\'e vectors} \label{App: Equal}
	
	In the case of uniaxial heterostrain (Sec. \ref{subsec:Uniaxial}), by solving the angle equation (\ref{eq:cosb}) for $\phi$ one can get the needed strain parameters to obtain equal length moir\'e vectors with an angle $\beta$ between them.  Taking into account the symmetrical solutions, we find 
	\begin{align}
		\epsilon_{s,r} & =\frac{4s}{1-\nu}\frac{f_{r}}{\sqrt{1-f_{r}^{2}}}\tan\left(\theta/2\right),\label{eq:betasolsE}\\
		\phi_{s,r} & =-\frac{s}{2}\arccos f_{r}+\frac{\pi}{3}\left(n+\frac{1}{2}\right),
		\label{eq:betasols}
	\end{align}
	where
	\begin{align}
		f_{r}\left(\nu,\cos\beta\right) & =\left(\frac{1-\nu}{1+\nu}\right)\frac{2+\cos\beta+r\sqrt{3}\left|\sin\beta\right|}{1+2\cos\beta}.
	\end{align}
	Here $s,r=\pm1$, and $n$ is an integer. The solutions are given in terms of four roots, which correspond to four equivalent strain directions that yield the same angle $\beta$. For both $r=\pm1$ one has two strain angles $\phi$ which are related by $\phi_{-,r}+\phi_{+,r}=\pi/3+n\pi$. Consequently there is always two strain angles, $\phi_{+}$ and $\phi_{-}=\pi/3-\phi_{+}$, with corresponding strain magnitudes $\pm\epsilon_{r}$, which give the same moir\'e pattern. Each angle $\phi_{\pm}$ is, in turn, symmetrical under the exchange $\phi_{\pm}\rightarrow\phi_{\pm}+\pi/3$, due to the honeycomb symmetry of the lattice. The $r=1$ roots correspond to the moir\'e patterns formed through the lateral contraction of the honeycomb lattices, as measured by the Possion's ratio, and thus correspond to larger strain magnitudes. While the $r=-1$ roots are solutions for any angle $\beta$, the roots $r=1$ are only solutions for certain $\beta$. The corresponding equal length of the moir\'e vectors reads
	\begin{align}
		\frac{\left|\mathbf{g}_{i}\right|^{2}}{\left|\mathbf{b}_{i}\right|^{2}} & =\frac{\left(1+\nu\right)^{2}f_{r}^{2}-\left(1-\nu^{2}\right)f_{r}+\left(1-\nu\right)^{2}}{\left(1-f_{r}^{2}\right)\left(1-\nu\right)^{2}}4\sin^{2}\left(\frac{\theta}{2}\right).
	\end{align}
	It is important to note that the strain angle $\phi$ is measured with respect to the orientation of the (non-deformed) honeycomb lattice. Upon rotation of both hexagonal monolayers by $\pm\theta/2$, the actual strain direction relative to each lattice is $\pm\theta/2+\phi$. Although the axis from which $\phi$ is measured depends on the chosen frame of reference (i.e., the lattice vectors $\mathbf{a}_{i}$), the actual direction of the strain, in relation to the orientation of the honeycomb primitive cell (hexagon), is always fixed.
	
	\section{Construction of the moir\'e Brillouin zone} \label{App: Const}
	
	\begin{figure}[t]
		\includegraphics[scale=0.32]{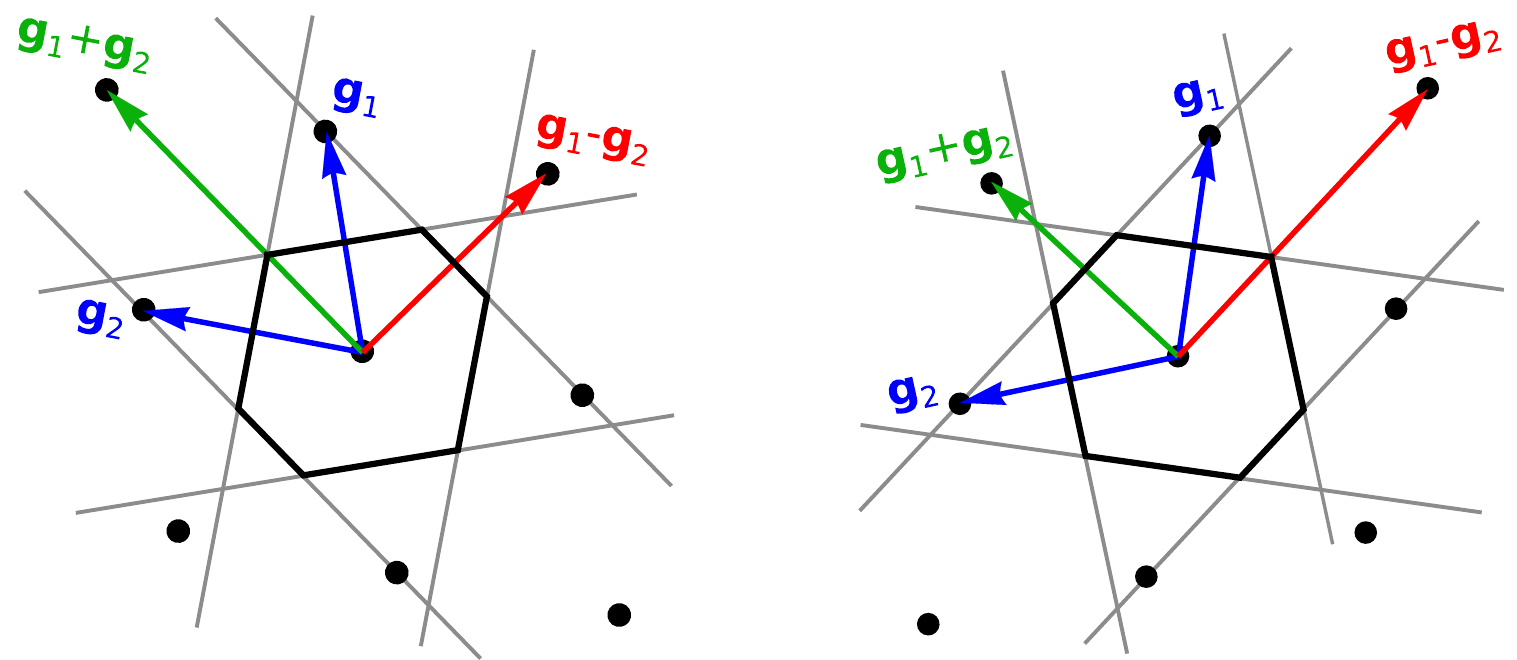}	
		\caption{Construction of the mBZ for equal length lattice vectors $\mathbf{g}_{1}$ and $\mathbf{g}_{2}$, with angles between them $\beta=70^{\circ}$ (left) and $\beta=110^{\circ}$ (right). The Bragg lines are shown in light gray, whose interceptions determine the mBZ (shown in black). If $\beta<90^{\circ}$, the interceptions are between the Bragg lines associated with the vectors $\mathbf{g}_{1}$, $\mathbf{g}_{2}$, $\mathbf{g}_{1}-\mathbf{g}_{2}$ (and their negatives), whereas if $\beta>90^{\circ}$ the interceptions are between the Bragg lines of $\mathbf{g}_{1}$, $\mathbf{g}_{2}$, $\mathbf{g}_{1}+\mathbf{g}_{2}$. Note that, up to a rotation, both cases have the same mBZ, since they represent the same lattice. The transition at which $\left|\mathbf{g}_{1}-\mathbf{g}_{2}\right|$ becomes larger (or smaller) than $\left|\mathbf{g}_{1}+\mathbf{g}_{2}\right|$ occurs at the critical square case $\beta=90^{\circ}$, where $\left|\mathbf{g}_{1}+\mathbf{g}_{2}\right|=\left|\mathbf{g}_{1}-\mathbf{g}_{2}\right|$
			and the six points of the mBZ are reduced to four. }
		\label{fig:BZconstruction}
	\end{figure}
	
	Consider two equal length vectors $\mathbf{g}_{1}$ and $\mathbf{g}_{2}$ with angle $\beta$ between them. We set, without loss of generality, the vector $\mathbf{g}_{1}$ on the $x$ axis, 
	\begin{align}
		\mathbf{g}_{1} & =g\left(1,0\right),\\
		\mathbf{g}_{2} & =g\left(\cos\beta,\sin\beta\right).
	\end{align}
	For any reciprocal vector $\mathbf{R}\left(m_{1},m_{2}\right)=m_{1}\mathbf{g}_{1}+m_{2}\mathbf{g}_{2}$, the corresponding Bragg line, which we shall denote as $l\left(m_{1},m_{2}\right)$, crosses $\mathbf{R}$ perpendicularly at $\mathbf{R}/2$. Since the mBZ has a mirror symmetry at $\beta=\pi/2$ by a reflection at the $x$ axis, it is sufficient to consider $\beta<\pi/2$. In that case the six intersections are between the set of two Bragg lines 
	\begin{align}
		l\left(1,0\right);\; & l\left(0,1\right),\\
		l\left(0,1\right);\; & l\left(-1,1\right),\\
		l\left(-1,0\right);\; & l\left(-1,1\right),
	\end{align}
	and their negatives (see Fig. \ref{fig:BZconstruction}). Now, for an arbitrary vector $\mathbf{f}=\left(f_{x},f_{y}\right)$ in the $xy$ plane, a perpendicular vector is $\mathbf{n}=\mathbf{e}_{z}\times\mathbf{f}=\left(-f_{y},f_{x}\right)$, whose angle with the $x$ axis is $\alpha=\arctan\left(n_{y}/n_{x}\right)$. A perpendicular line to $\mathbf{f}$ that crosses $\mathbf{f}/2$ then reads $y=\left(f_{x}/f_{y}\right)\left(f_{x}-x\right)+f_{y}$. Therefore, since $\mathbf{g}_{1}$ is all in $x$, the three lines that we need for the mBZ construction are
	\begin{align}
		l\left(1,0\right) & : x_{1}=\frac{g}{2},\\
		l\left(0,1\right) & : y_{2}=\frac{1}{\tan\beta}\left(\frac{g}{2}\cos\beta-x\right)+\frac{g}{2}\sin\beta,\\
		l\left(-1,1\right) & : y_{3}=\frac{\cos\beta-1}{\sin\beta}\left(g\frac{\cos\beta-1}{2}-x\right)+\frac{g}{2}\sin\beta.
	\end{align}
	This leads to the three intersections points
	\begin{align}
		\mathbf{I}_{1} & =\frac{g}{2}\left[1,\frac{1}{\tan\beta}\left(\cos\beta-1\right)+\sin\beta\right],\\
		\mathbf{I}_{2} & =\frac{g}{2}\left[2\cos\beta-1,-\frac{1}{\tan\beta}\left(\cos\beta-1\right)+\sin\beta\right],\\
		\mathbf{I}_{3} & =\frac{g}{2}\left[-1,\frac{1}{\tan\beta}\left(\cos\beta-1\right)+\sin\beta\right].
	\end{align}
	We can write these points in terms of the vectors $\mathbf{g}_{i}$ as
	\begin{align}
		\mathbf{I}_{1} & =\frac{\mathbf{g}_{1}+\mathbf{g}_{2}}{2}\left(1+\frac{\mathbf{g}_{1}\cdot\mathbf{g}_{2}}{\mathbf{g}_{1}\cdot\mathbf{g}_{1}}\right)^{-1},\\
		\mathbf{I}_{2} & =-\mathbf{I}_{1}+\mathbf{g}_{2},\\
		\mathbf{I}_{3} & =\mathbf{I}_{1}-\mathbf{g}_{1}.
	\end{align}
	The case $\beta>\pi/2$ is obtained by a mirror reflection of $\mathbf{g}_{2}$ around $\mathbf{g}_{1}$, thus leading to Eq. (\ref{eq:Q}) after changing the notation of the interception points.

	\section{Electronic properties without strain fields} \label{sec:NoGauge}
	
	\begin{figure}[h]
		\includegraphics[scale=0.70]{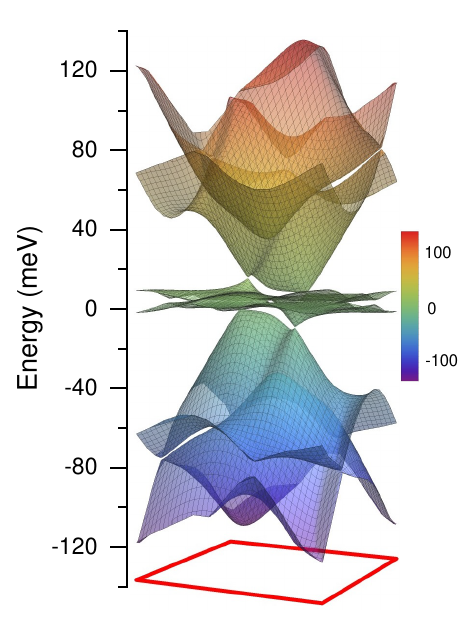}	
		\caption{Electronic structure of TBG with $\mathbf{A}=0$ and $V=0$. Other parameters are the same as in Fig.~\ref{fig:BandStructure}c).} 
		\label{fig:FigNoGauge}
	\end{figure}
	
	Figure~\ref{fig:FigNoGauge} shows the electronic structure of TBG under uniaxial heterostrain, but with zero gauge and scalar strain fields.  Parameters are the same as in Fig.~\ref{fig:BandStructure}c). Even in absence of gauge fields there is a distortion of the energy bands. As the strain increases, the mBZ is distorted, the Dirac cones are shifted and the remote bands are pushed to a region close to the narrow bands.

\end{document}